\documentclass[sigconf]{acmart}

\AtBeginDocument{%
  \providecommand\BibTeX{{%
    \normalfont B\kern-0.5em{\scshape i\kern-0.25em b}\kern-0.8em\TeX}}}

\setcopyright{none}
\renewcommand\footnotetextcopyrightpermission[1]{}
\acmConference{}{}{}
\renewcommand{\shortauthors}{}
	
\usepackage{float}

\usepackage{subcaption}

\usepackage{graphicx}
\usepackage{caption}

\usepackage{url}
\usepackage{hyperref}
\usepackage[hyphenbreaks]{breakurl}

\usepackage[ruled,vlined]{algorithm2e}
\usepackage{array,etoolbox}

\usepackage{booktabs}

\usepackage{amsmath}

\usepackage{epsfig,amstext}

\usepackage{enumitem}
\usepackage{multirow}
\usepackage{makecell}
\usepackage{subcaption}

\begin{document}

\title{Did State--sponsored Trolls Shape the 2016 US Presidential Election Discourse? Quantifying Influence on Twitter}


	\author{Nikos Salamanos}
	\affiliation{%
		\institution{Cyprus University of Technology}
		\streetaddress{}
		\city{}
		\country{}}
	\email{nik.salaman@cut.ac.cy}
	
	\author{Michael J. Jensen}
	\affiliation{%
		\institution{University of Canberra}
		\streetaddress{}
		\city{}
		\country{}}
	\email{Michael.Jensen@canberra.edu.au}
	
	\author{Costas Iordanou}
	\affiliation{%
		\institution{Cyprus University of Technology}
		\streetaddress{}
		\city{}
		\state{}
		\country{}}
	\email{costas.iordanou@eecei.cut.ac.cy }
	
	\author{Michael Sirivianos}
	\affiliation{%
		\institution{Cyprus University of Technology}
		\streetaddress{}
		\city{}
		\state{}
		\country{}}
	\email{michael.sirivianos@cut.ac.cy}
\renewcommand{\shortauthors}{Salamanos et al.}

\begin{abstract}
It is a widely accepted fact that state--sponsored Twitter accounts operated during the 2016 US presidential election, spreading millions of tweets with misinformation and inflammatory political content. Whether these social media campaigns of the so--called ``troll'' accounts were able to manipulate public opinion is still in question. Here, we quantify the influence of troll accounts on Twitter by analyzing 152.5 million tweets (by 9.9 million users) from that period. The data contain original tweets from 822 troll accounts identified as such by Twitter itself. We construct and analyse a very large interaction graph of 9.3 million nodes and 169.9 million edges using graph analysis techniques, along with a game--theoretic centrality measure. Then, we quantify the influence of all Twitter accounts on the overall information exchange as is defined by the retweet cascades. We provide a global influence ranking of all Twitter accounts and we find that one troll account appears in the top--100 and four in the top--1000. This combined with other findings presented in this paper constitute evidence that the driving force of virality and influence in the network came from regular users -- users who have not been classified as trolls by Twitter. On the other hand, we find that on average, troll accounts were tens of times more influential than regular users were. Moreover, 23\% and 22\% of regular accounts in the top--100 and top--1000 respectively, have now been suspended by Twitter. This raises questions about their authenticity and practices during the 2016 US presidential election.
\end{abstract}


\maketitle

\section{Introduction} 
\label{sec:intro}

The Russian efforts to manipulate the outcome of the 2016 US presidential election were unprecedented in terms of the size and scope of the operation. Millions of posts across multiple social media platforms gave rise to hundreds of millions of impressions targeting specific segments of the population in an effort to mobilize, suppress, or shift votes~\cite{Jamieson:2018}. Trolls were particularly focused on the promotion of identity narratives~\cite{Jensen:2018}, though that does not distinguish them from many other actors during the election \cite{Sides:2018}. The Special Counsel's report described this interference as ``sweeping and systematic''~\cite[vol 1, 1]{Mueller:2019}. Russian efforts focused on inflicting significant damage to the integrity of the communication spaces where Americans became informed and discussed their political choices during the election \cite{Mazarr:2019}. Therefore, the question of whether these disinformation campaigns had a significantly real impact on social media is of paramount importance. Although there is a large body of work that tried to address this question \cite{Benkler:2018,Jamieson:2018,Sides:2018}, a conclusive result is still elusive. 

In this paper, we address this question by measuring the influence of the so--called ``troll'' accounts together the virality of information that they spread on Twitter during the period of 2016 US Presidential election. Let us note that ``troll'' is any account that deliberately spreads disinformation, tries to inflict conflict or causes extreme emotional reactions. A troll account could be human or operated automatically. An automated operated account is called ``bot''and is controlled by an algorithm that autonomously performs actions on Twitter. The term ``bot'' is not synonymous to ``troll'' as benign bots do operate and have positive impact on users\footnote{\url{https://blog.mozilla.org/internetcitizen/2018/01/19/10-twitter-bots-actually-make-internet-better-place/}}. In fact, Twitter has set specific rules for acceptable automated behavior\footnote{\url{https://help.twitter.com/en/rules-and-policies/twitter-automation}}.

There are several obstacles for any empirical study on this subject: (i) the lack of complete and unbiased Twitter data -- the Twitter API returns only a small sample of the users' daily activity; (ii) Tweets from deactivated profiles are not available; (iii) The followers and followees lists are not always accessible, hence the social graph is unknown. 

Having that in mind, the rationale of this study is the following: First, we collected 152.5 million election--related tweets during the period of 2016 US presidential election, using the Twitter API along with a set of track terms related to political content. The data contain original troll tweets from that period which later on were deleted by Twitter. Secondly, based on ground--truth data released by Twitter itself, regarding state--sponsored accounts linked to Russia, Iran, Venezuela and Bangladesh states, we identified 822 trolls present in our data. Then, we constructed a very large \textit{interaction--graph} of 9.3 million nodes/users and 169.9 million edges. Using graph analysis techniques and Shapley Value--based centrality we analyze (i) the graph structure; (ii) the diffusion of potentially political content as represented by the retweet cascades of tweets with at least one web or media URL embedded in the text. 

Our approach is agnostic with respect to the actual political content of the tweets. The goal is to measure the impact of all users on the overall diffusion of information and consequently estimate the impact of ground--truth trolls. For the rest of the paper, we call \textit{``regular''} the users that have not been classified as trolls by Twitter; they are just the rest of the population and might not always represent benign accounts.

\noindent \textbf{Research Questions (RQ):} We address the following RQ:
\begin{itemize}
\item[\textbf{RQ1:}] Who are the most influential trolls and regular users? Can we rank them in order of contribution (impact) to the overall information diffusion?
\item[\textbf{RQ2:}] Which are the viral retweet cascades initiated by regular users and specific troll accounts?
\item[\textbf{RQ3:}] What is the proximity of top--k influential regular users to bot accounts and how many of them have been suspended by Twitter, later on?
\end{itemize}

\noindent \textbf{Contributions:} Our primary contributions are as follows:
\begin{enumerate}
\item We construct one of the largest graphs representing the interactions between state--sponsored troll accounts and regular users in Twitter during the 2016 US Presidential election. This counts as an approximation of the original social graph.
\item We introduce the notion of \textit{flow graphs} -- a natural representation of the information diffusion that takes place in Twitter platform during the retweet process. This formulation allows us to apply a Shapley Value--based centrality measure for a fair estimation of users' contribution on the information shared, without imposing assumptions on the users' behavior. Moreover, we estimate the virality of retweet cascades by the \textit{structural virality} along with the influence each user has on them by the \textit{influence--degree}.
\item By answering the research questions, we present strong evidence that troll activity was not the main cause of viral cascades of web and media URLs in Twitter. Our measurements show that the regular users were in general the most active and influential part of the population and their activity was the driving force of the viral cascades. At the same time, we find that on average, trolls were tens of times more influential than regular users -- an indicator of the effectiveness of their strategies to attract attention. These findings further substantiate previously reported insights~\cite{Vosoughi:2018,Zannettou:www2019,Zannettou:WebSci2019}. Furthermore, more than 20\% of the top--100 as well as the top--1000 regular users have now been suspended by Twitter. This sets their authenticity in question as well as their activity during that period.
\end{enumerate}

\noindent\textbf{Data availability:} Part of the dataset is available under proper restrictions for compliance with Twitter's ToS and the GDPR\footnote{\url{https://drive.google.com/drive/folders/1IglMNUNqFzC5ndKXfbdMgDCKgsf1mPE_?usp=sharing}}. The ground truth data are provided by Twitter\footnote{\label{twitter}\url{https://about.twitter.com/en/our-priorities/civic-integrity}}.

\section{Related Work}
\label{sec:literature} 

\begin{table*}[!t]
    	\setlength\tabcolsep{8pt}
	\caption{Ground--truth Twitter data}
	\label{tab:Twitter_groundtruth}
	\centering
	\begin{tabular}{l | c c || c || c  c}
		\multicolumn{2}{c}{}  &     & \ Account creation date $<$ 08/11/2016 & \multicolumn{2}{c}{21/09/2016 -- 07/11/2016}\\
		  &   \#Trolls      & \#Trolls tweeted  &      \#Trolls     &  \#Trolls tweeted     & \#Trolls in our data \\	    
		\hline		
		Russia          & 4,024     & 3,838       & 3,716                        & 527                        &  294        \\
		Iran            & 3,081     & 2,861       & 1,399                        & 1,041                      &  426        \\
		Venezuela      & 1,951     & 1,565       & 693                          & 414                        &  101        \\
		Bangladesh       & 15        & 11          & 6                            & 3                          &   1         \\
		\hline	 
		Total accounts  & 9,071     & 8,275       & 5,814                        & 1,985                      &  822     \\
		Total tweets    & N/A       & 25,076,853  & 21,854,780                   & 843,789                    &  657,979    \\
		\hline
	\end{tabular}	                                  
\end{table*}

In a seminal work on the general problem of disinformation on Twitter~\cite{Vosoughi:2018}, the authors investigated the diffusion cascades of true and false rumors disseminated from 2006 to 2017 -- approximately 126K rumor cascades spread by 3 million people. The main findings are (i) false news diffused faster and more broadly than the true ones; (ii) human behavior contributes more to the spread of falsity than the trolls. These findings are in line with our main result, that is, the regular users had the dominant role on the viral cascades. Moreover, part of our methodology on the construction of the retweet trees has been inspired by this work. Since our main goal is to quantify the impact of users on the overall information exchange, we do not classify the tweets content as fake and non fake rumors. We use the URLs that have been spread by trolls to serve as ``anchors'' of retweet cascades that contain the same piece of information with the troll tweets.

In~\cite{Bovet:2019} the authors analyzed 171 million tweets by 11 million users -- collected five months prior to the 2016 US presidential election. Using these data, they examined 30 million tweets which contained at least one web URL pointing to a news outlet website. 25\% of these news were either fake or biased representing the spreading of misinformation. Then, they investigated the flow of information by constructing retweet networks for each news category. Furthermore, they estimated the most influential spreaders in the retweet networks using the Collective Influence (CI) algorithm~\cite{Morone2015}. One of their findings is that the Trump supporters were the main group of users that spread fake news although it was not the dominant one in the whole network. We note that in~\cite{Bovet:2019}, the overall retweet graph is directly constructed by the data as they were provided by the Twitter API. In our study, we enrich the raw Twitter data by considering all the possible information paths and at the same time we provide an estimation of the retweet trees. 

Grinberg et al.~\cite{Grinberg2019} investigates the extent to which Twitter users were exposed to fake news during the 2016 US presidential election. Their data consists of tweets from 16.4K Twitter accounts that were active during the 2016 US election season, along with their list of followers. They restrict their analysis on tweets containing a URL from a website outside the Twitter. Moreover, the authors introduce the notion of users' ``exposures'', i.e., tweets from a user to his followers. This approach is roughly in line with the \textit{flow graphs} that we present in Section~\ref{subsec:flow-graph-tree}. One of their main findings is that although a large part of the population had been exposed to fake news, only a small fraction (1\%) was responsible for the diffusion of 80\% of fake news.

In~\cite{Zannettou:www2019,Zannettou:WebSci2019} the authors analyzed the characteristics and strategies of 5.5K Russian and Iranian troll accounts in Twitter and Reddit. Moreover, using \textit{Hawkes Processes} they compute an overall statistical measure of influence that quantifies the effect these accounts had on social media platforms, such as Twitter, Reddit, 4chan and Gab. One of their main results is that even though the troll accounts reach a considerably large number of Twitter users and are effective on spreading URLs on Twitter, however, their overall effect on the social platforms is not dominant. Our findings verify these results and support the fact that some trolls have above average influence. 

In~\cite{Badawy:ASONAM2018, Badawy:2019} the authors examined the Russian disinformation campaigns on Twitter in 2016, based on 43M tweets shared by 5.7M users and 221 trolls. They focused on the characteristics of \textit{spreaders}, namely the users that had been exposed and shared content previously published by Russian trolls. They constructed the retweet graph by mapping retweet actions to edges. Then, they applied the label propagation algorithm in order to classify Twitter accounts as either conservative or liberal. Finally, they used \textit{Botometer}~\cite{Davis:2016}, in order to determine whether spreaders and non--spreaders can be labeled as bots. We also apply this technique in order to examine whether the top--k influential users exhibit bot behavior.

In~\cite{Huyen:ASONAM2019} a postmortem analysis is conducted in one million Twitter accounts which although were active during the 2016 US election period, later on they were suspended by Twitter. The authors focused on the community level activities of the suspended accounts and for that purpose, they clustered them into communities. Then, they compared the characteristics of suspended account communities with the not suspended ones and they found significant differences in their characteristics, especially in their posting behavior.

Bovet et al.~\cite{Bovet2018} developed a method to infer the political opinion of Twitter users during the 2016 US presidential election. For that purpose, they constructed a directed social graph based on the users' actions (replies, mentions, retweets) between them -- a similar graph formulation technique to ours, in this paper. Then, they monitored the evolution of three structural graph properties, the Strongly Connected Giant Component, Weakly Connected Giant Component, and the Corona.

\begin{table}[!t]
	\centering
	\caption{Our Twitter dataset}
	\label{tab:our_dataset}
	\setlength\tabcolsep{9pt}
	\begin{tabular}{l|l|l}
		&    Regular users  & Trolls   \\         
		\hline		
		Unique Twitter accounts     &  9,939,698     & 822    \\
		Total tweets         &  152,479,440   & 35,489  \\
		\hline\hline
		Replies              &  12,942,628    & 129     \\
		Mentions             &  172,145,775   & 33,627  \\	 
		Retweets             &   N/A         & N/A      \\	
		\hline
		\multicolumn{3}{l}{\parbox{7.6cm}{\textbf{Track terms:} 'ben carson', 'bencarson', 'bernie sanders', 'bernie2016', 'bettercandidatethanhillary', 'carlyfiorina2016', 'carson2016', 'clinton', 'clinton2016', 'cruz2016', 'cruzcrew', 'cruzintocaucus', 'donaldtrump', 'donaldtrump2016', 'dumptrump', 'election2016', 'feelthebern', 'fiorina', 'fiorina2016', 'fitn', 'heswithher', 'hilaryclinton', 'hillary2016', 'hillaryclinton', 'hrc', 'huckabee', 'huckabee2016', 'imwithher', 'iwearebernie', 'jill stein', 'jillstein', 'johnkasich', 'kasich', 'kasich2016', 'kasich4us', 'letsmakeamericagreatagain', 'makeamericagreatagain', 'makeamericawhiteagain', 'marco rubio', 'marcorubio', 'martinomally', 'mikehuckabee', 'nevertrump', 'newyorkvalues', 'nhpolitics', 'nhpsc', 'omalley', 'paul2016', 'primaryday', 'randpaul2016', 'readldonaldtrump', 'realdonaldtrump', 'redstate', 'rick santorum', 'ricksantorum', 'rubio2016', 'rubiowa', 'sensanders', 'sentedcruz', 'stein2016', 'teamKasich', 'teamcarly', 'teamclinton', 'teamcruz', 'teamhillary', 'teammarco', 'teamrubio', 'teamtrump', 'ted cruz', 'tedcruz', 'the donald', 'thedonald', 'therealdonaldtrump', 'trump', 'trump2016', 'trumptrain', 'unitedblue'}}\\
		\hline
	\end{tabular}	                                  
\end{table}

\section{Datasets} 
\label{sec:data} 

\textbf{Ground--truth Twitter data:} Twitter has released a large collection of state--sponsored trolls activities as part of Twitter's election integrity efforts (see footnote 4). This is an ongoing work, where the list of malicious accounts are constantly updated. We requested the unhashed version which consists of 9,071 trolls accounts information affiliated with Russia, Iran, Venezuela and Bangladesh states, along with 25M tweets shared by 8,275 trolls (Table~\ref{tab:Twitter_groundtruth}). The number of accounts represented in the tweets is less than the total number of accounts listed as trolls (which is noted by Twitter in the description document of the data). In this study, we leverage only the troll IDs which served as ground--truth identifiers of the trolls in our tweets collection we present next.

\textbf{Our Twitter dataset:} Our analysis is based on 152.5M tweets from 9.9M users. We downloaded the data using the Twitter streaming (1\%) and Tweepy\footnote{\url{https://www.tweepy.org/}} Python library, in the period before and up to the 2016 US presidential election -- from September 21 to November 7, 2016 (47 days; we did not collect data on 02/10/2016). The tweets' track terms\footnote{\url{https://developer.twitter.com/en/docs/tweets/filter-realtime/guides/basic-stream-parameters.html}} were related to political content such as ``hillary2016'', ``clinton2016'', ``trump2016'' and ``donaldtrump2016'' -- namely, a list of phrases used to determine which Tweets are delivered by the stream (see Table~\ref{tab:our_dataset}). In addition to the tweet text, user screen name, and user ID, we also collected metadata including the hashtags, the URLs and mentions that were included in the tweet text, as well as information on the account creation, user timezone, and user location. Based on the ground--truth Twitter data, we identified 35.5K tweets from 822 troll accounts (Table~\ref{tab:our_dataset}). This dataset was originally collected for a different project that did not need the retweet labels, which, therefore, were not collected. Still, as we describe in Section~\ref{subsec:RT_cascades}, we were able to recover a large portion of the retweets based on the JSON object information that we have already collected.

The question is whether the collected tweets are a representative sample of the trolls' activities during the 2016 US presidential election, since only 10\% of the listed ground--truth trolls are present. Therefore, we proceed with a general evaluation of our dataset based on the trolls' account information along with the trolls' tweets provided by Twitter. First, we find those trolls with account--creation date before November 8, 2016 -- since November 7, 2016 was the last day of our crawling. Then, we list the trolls that have an activity between the dates we collected our data (Table~\ref{tab:Twitter_groundtruth}). 5,814 out of 9,071 trolls joined Twitter before 08/11/2016 and 1,985 of them have actively tweeted during the period 09/21/2016 -- 07/11/2016, with 844K tweets. At the same time, 822 out of the 1,985 trolls (41\%) are present in our data. These accounts were responsible for the 78\% of the total trolls' activity (658K tweets), during that period. In summary, the data we collected contain the 41\% of the active trolls and some of them were the most active ones.

\section{Methodology}
\label{sec:methods} 

\subsection{The Social network}
\label{subsec:graph}

We leverage the users' activity as is recorded in the data (152.5M tweets) to construct an approximation of the
follower--graph -- the true social network which is not publicly available to a large extent. Specifically:

\textbf{Interaction--graph:} In short, we map users to nodes and interactions between users to edges. In Twitter, the interactions between users belong to three categories: (i)
\textit{replies}; (ii) \textit{retweets} or \textit{quotes} -- a special form of retweet; (iii) \textit{mentions}. We define the directed edge $(i, j)$, from user $i$ to user $j$, for every action of $i$ on tweets of $j$. For example, if $i$ had replied to a tweet of $j$. The direction of the edge implies that $i$ is a \textit{follower} of $j$, while the reverse direction represents the information flow from $j$ to $i$. 

This process outputs a directed \textit{multigraph}, where many edges may connect the same pair of users. It consists of 169,921,912 edges, 9,321,061 regular users and 821 trolls. Although we did not collect the original retweet labels, this information still appears in the field ``mentions'' provided by the Twitter API. Hence, the retweet actions are included in the constructed graph. Even though the number of troll accounts is small, there are some indications that some troll accounts might have substantial activity which is worth for further investigation. For instance, we have 671K edges (i.e. interactions) that point to 285 trolls; in other words, more than half million users had an interaction with troll accounts. Moreover, users who although tweeted neither performed an action to other accounts, nor received actions from others, they cannot be part of the graph. That is why, the total number of nodes is not equal with the total number of users who appear in the initial dataset. 

\textbf{Follower--graph:} Finally, we construct the follower--graph by discarding the duplicate edges and keeping only the earliest one. Obviously, it is a directed graph with 9.32M users/nodes and 84,1M edges and represents an approximation of the true follower--graph. 

\subsection{Retweet Cascades}
\label{subsec:RT_cascades}

\begin{figure*}[!t]
	\centering
	\includegraphics[width= .90\linewidth]{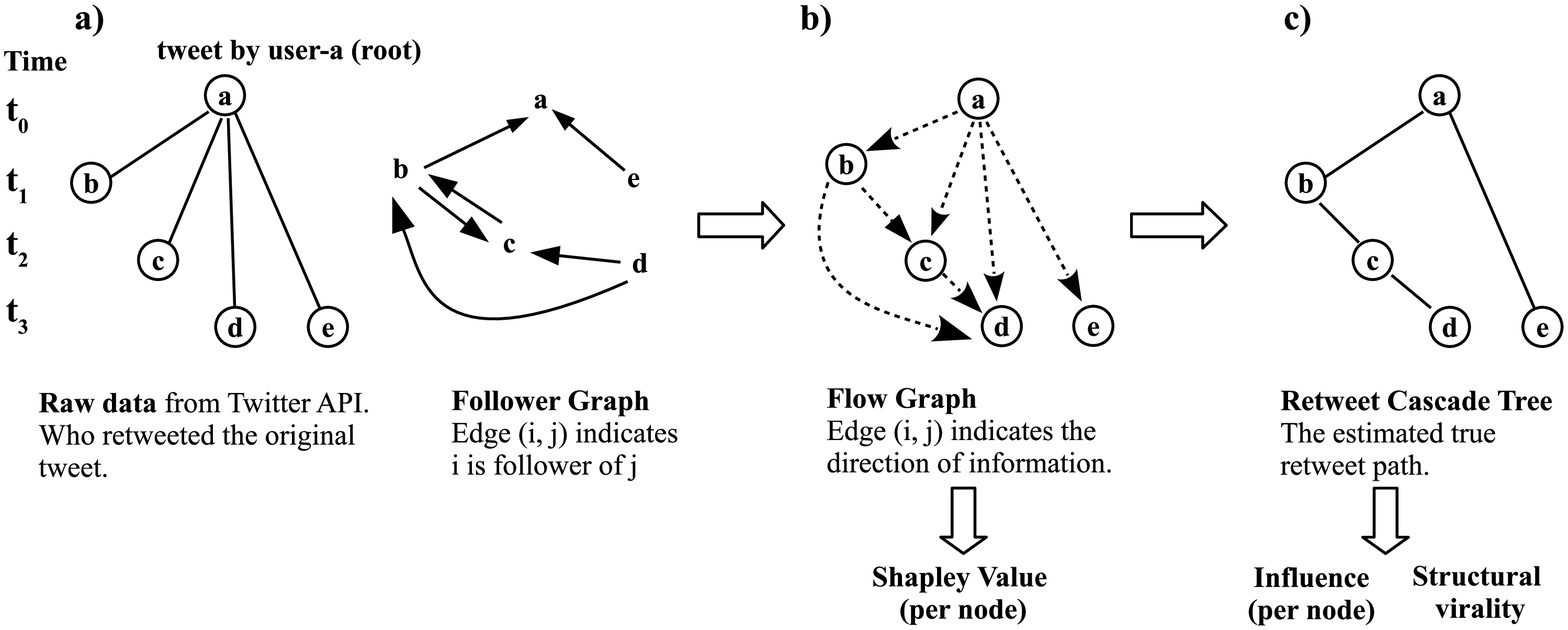} 
	\caption{Toy example of retweet analysis. (a) The raw data provided by Twitter API along with the follower--graph. (b) The flow graph shows the full information flow according to Twitter functionality and the follower--graph. The edges present the path of information that appears on the users' timeline prior to their retweets. For instance, user $c$ has retweeted on date $t_{2}$. At the same time user $b$, whom user $c$ follows, has retweeted on date $t_{1}<t_{2}$. Note that a given retweet contains both the name of the user who retweeted and the name of the root user who posted the original tweet. Hence, we have an edge from the root to any retweeter because the users have retweeted the root tweet even if they did not follow the root user. (c) The time-inferred cascade tree is constructed from the flow graph by making the assumption (see Section \ref{subsec:flow-graph-tree}), that each retweeter has been influenced by the friend who just recently retweeted the original tweet.}
	\label{fig:example_graph}
\end{figure*}

When a user retweets, usually he/she agrees with the context of the original tweet (root--tweet) that has been retweeted. For this reason, the analysis of the retweet cascades -- i.e. a series of retweets upon the same root--tweet -- is important for the identification of the viral cascades as well as the influential users in them. Hence, we have to recover the retweet labels that are missing from the data. We are able to achieve that by leveraging the following facts regarding the retweets: (1) The beginning of a retweet text always has the form ``RT @$\langle$user screen name$\rangle$:'' where the $\langle$user screen name$\rangle$ refers to the author (root--user) of the original tweet; (2) The ``Entities'' JSON section includes the ``mentions'' -- the list of user IDs along with their screen names for all users mentioned in the tweet. Hence, in ``mentions'' there is always listed the root--user ID and his/her screen name; (3) In addition, the ``Entities'' field provides the list of URLs that are embedded in the tweet text; web or media URL (i.e. media material such as videos and photos). 

\begin{table}[!t]
	\centering
	\caption{Retweet cascades with minimum 100 unique retweeters}\label{tab:retweets}
	\setlength\tabcolsep{18pt}
	\begin{tabular}{l | l | l}
		& Regular Users     & Trolls \\
		\hline
		Total users                 & 3,633,457 & 233                 \\
		\hline
		Root users                  & 8,192     & 12                  \\
		Root tweets                 & 45,986    & 423                 \\
		\hline			
		Retweeters                  & 3,630,764	& 228                 \\	
		\hline		
		Total retweets   & \multicolumn{2}{|c}{19,588,072}            \\
		Total URLs       & \multicolumn{2}{|c}{43,989}                \\                
		\hline
	\end{tabular}	                                  
\end{table}

Based on these observations, we identify a set of tweets as retweets to the same root--tweet doing the following: (1) We collect those tweets for which the head of the text has the form ``RT @$\langle$user screen name$\rangle$:'' and at least one URL is embedded in the text. (2) We match the $\langle$user screen name$\rangle$ with the corresponding user ID (by the ``mentions'' JSON field) identifying the root--user ID. (3) We group these filtered tweets by the tweet--text plus the root--user ID. Finally, we filter these groups further by the condition that at least 100 distinct retweeters have tweeted the root--tweet (excluding the root--user since he/she may have retweeted his/her own tweet). 

We reconstruct only the retweet cascades where the root tweet--text contains at least one URL. The reason for that is that the URLs have a dual role. First, they are strong identifiers of the tweet--text equality by which we reconstruct the retweet cascades. Secondly, in a retweet cascade, it is not only the actual tweet that has been diffused, but mainly the information it contains. So, the URLs serve as ``anchors'' by which we connect distinct retweet cascades, considering that they are referring to the same piece of information (see also Section~\ref{subsubsec:results_shapley}). Finally, let us note that the term ``RT'' is not an absolute indication by itself, that a tweet is a retweet\footnote{\url{https://help.twitter.com/en/using-twitter/retweet-faqs}}. A user could have just typed ``RT'' at the beginning of his/her tweet--text. We exclude this exception by the condition of at least 100 repetitions of the tweet--text. 

In conclusion, this process enables us to recover 46.4K retweet cascades consisting of 19.6M tweets, (Table~\ref{tab:retweets}). In order to verify the accuracy of this approach, we recollect the tweets that we have identified as retweets by the previous method. Only 10M tweets out of 19.6M are still available while the rest have been deleted. Then, in these 10M tweets, we verified that indeed our mapping of retweets to root--user IDs is always correct. Nevertheless, there were some discrepancies in only 9,689 tweets (out of the 10M recollected tweets), regarding the mapping of retweets to root--tweet IDs. There are two cases that our approach is not able to capture: (i) The root--user could post the same tweet more than once. In this case, although the tweet text and all the other information are identical across the retweets, the root--tweet IDs are different; (ii) A given user might quote instead of just retweet. In this case, the quoted retweet will differ in the text compared to the other retweets, even though the root--tweet IDs are identical. 

\subsection{Retweet cascade tree and Flow graph}
\label{subsec:flow-graph-tree}

Generally, the retweet data returned by the Twitter API have by design limited information regarding the true chain of retweet events. For a given retweet, the information provided is the retweeter ID and the root--user ID. Hence, in terms of influence, this corresponds to the case where all the retweeters have been influenced by the root--user. In Figure~\ref{fig:example_graph}(a) we present an example of the raw data. This star--like cascade structure does not depict the true chain of retweet events since a user may have retweeted a retweet of a friend.

\textbf{Retweet cascade tree:} A widely used method for the reconstruction of the true retweet path is the time--inferred diffusion process~\cite{Goel2015:virality,Vosoughi:2017,Vosoughi:2018}. It is based on the causality assumption that a given user before retweeting has been influenced by his ``friend'' who has recently retweeted the same original tweet. Moreover, since a user can retweet a tweet more than once, we assume that he has been influenced by another user on his first action only. Hence, the final retweet path (see Figure~\ref{fig:example_graph}(c)) is constructed by the raw data provided by Twitter in conjunction with the follower--graph (Figure~\ref{fig:example_graph}(a)). Thus, we have two rather extreme cases, the one is the star tree that we take from Twitter API where no real diffusion structure is present, and the other is the cascade tree where a specific hypothesis has been applied with respect to who was influenced by whom. The latter emphasizes the most recent friend whereas the former always the root user. In order to define an intermediate case, we introduce the notion of \textit{flow graph}.

 \textbf{Flow graph:} The flow graph presents the direction of all possible influence between the retweeters that may have taken place by the information--diffusion in the Twitter platform. Let us consider the toy example in Figure~\ref{fig:example_graph}. Before constructing the retweet cascade tree in Figure~\ref{fig:example_graph}(c), we first have to identify all the time-inferred edges from the users that retweeted in time $t$ to the users who will retweet in $t+1$. The edges direction indicates the information flow on the Twitter platform and is based on the fact that when a user retweets a given tweet, his action appears on his followers' timeline. For instance, when user $b$ retweets the root tweet in $t_{1}$, he is transmitting this information to his followers $c$ and $d$. Finally, we add an edge from the root user to any of the retweeters because in any given retweet, the author's screen name is always visible. The construction of the flow graphs is based on the follower--graph, where the edges are time inferred. So, in a given time $t_{i}$, a given user $i$ receives information from the users he had already started following at a certain time $t< t_{i}$. 

The flow graph together with the retweet tree are the two graph structures we leverage in order to evaluate the impact of users in the overall information exchange. In summary: 

\noindent\textbf{Flow graph:} We measure the contribution of the users to the overall diffusion of information by the Shapley Value--based centrality. \textbf{Retweet cascade tree:} We measure the influence of every user in a given retweet tree by the influence--degree, and the overall virality of the tree by the structural virality.

\subsection{Shapley Value--based centrality}
\label{subsec:shapley} 

Towards evaluating the users in terms of the influence/impact they had on the retweet cascades we have to create a consistent ranking where the top--k users are the most influential ones. One way to do so, is to use a centrality measure that fits well in our problem. Here, we apply the Shapley Value--based degree centrality~\cite{ShapleyDiscounted:2014,shapley:journal:2014,shapley:wine:2010} one of the game--theory inspired methods of identifying influential nodes in networks~\cite{Narayanam:2008,Narayanam:2011,Manilla:2011}. These methods are based on the Shapley Value~\cite{Shapley:1953}, a division scheme for fair distribution of gains or costs in each player of a cooperative game. The Shapley Value of each player in the game is the average weighted marginal contribution of the player over all possible coalitions. Hence, the problem of computing the Shapley Value in a $N$ player game has in most cases exponential complexity, since the possible coalitions are $2^{N}$.

We apply the Shapley Value--based degree centrality introduced in~\cite{shapley:journal:2014,shapley:wine:2010} which is further refined in~\cite{ShapleyDiscounted:2014}. First, in~\cite{shapley:journal:2014,shapley:wine:2010} the authors provide a linear time algorithm for the exact computation of the Shapley Value in the following game. Given a directed graph $G(V, E)$, with $V$ nodes and $E$ edges, the set of players are the nodes in $V$ and each coalition is a subset of $V$. The value of a coalition $C$ is defined by the size of the set $fringe(C)$ i.e. the set that consists of the members of $C$ along with their out--neighbors. This set represents the sphere of influence of the coalition $C$. Moreover, we define that the value of the empty coalition is always zero. The exact closed form solution of the Shapley Value of a node $u_{i}$ is \\
$SV(u_{i}) = \displaystyle\sum_{u_{j} \in \{u_{i}\} \cup N_{out}(u_{i})}\frac{1}{1 + \text{indegree}_{G}(u_{j})}$.\\ Hence, the algorithm for computing the Shapley Values has running time $O(|V | + |E|)$ (see Algorithm 1 in~\cite{shapley:journal:2014,shapley:wine:2010}). In fact, the Shapley Value is the sum of probabilities that the node contributes to each of its neighbors and also itself. 

This formulation is very similar to what we want to measure in the flow graphs. In our case, the value of a coalition is the set of users that have been informed from the members of the coalition about a given root--tweet. Having said that, we cannot directly apply the above formulation since a node cannot inform itself. This very problem has been addressed by the authors in~\cite{ShapleyDiscounted:2014} to solve the influence maximization problem. They refined the previous formulation so that the value of a coalition $C$ is the size of the out--neighbors of the member in $C$, i.e. the number of nodes that can be directly influenced by $C$.
In conclusion, we compute the Shapley Value for all nodes in any flow graph using the following formula (see~\cite{ShapleyDiscounted:2014}): 
\begin{equation}\label{eq:shapley}
	SV(u_{i}) = \sum_{u_{j} \in N_{out}(u_{i})}\frac{1}{1 + \text{indegree}_{G}(u_{j})}
\end{equation}
In this way, the ``leaf'' nodes always have zero Shapley Value since they did not inform anyone in the flow graph. The advantage of this approach is that it provides a linear time computation of Shapley Values and also works for disconnected graphs. In fact, this is the case that we face here, since the overall information flow is represented by the flow graphs i.e. a set of disjoint graphs. Moreover, we can compute the overall Shapley Value for any subset of retweet cascades that a user is part of. As we will show in Section~\ref{subsubsec:results_shapley} we use this property in order to evaluate trolls and regular users together, only in a subset of retweet cascades.
The intuition of this approach is that Equation~\ref{eq:shapley} computes in a fair way the users' contribution in informing the other members of the graph for a given piece of information, which in our case is the original root--tweet and the URL it contains. We note that from the method in~\cite{ShapleyDiscounted:2014} we use only the part that computes the Shapley Values and not the whole process (influence maximization). Our goal is to compute the users' contribution without assumptions regarding the influence process.

Finally, the global Shapley Value of a user in the overall information exchange is the summation of his Shapley Values in the flow graphs (FG) the user participates in. Hence: 
\begin{equation}\label{eq:global_shapley}
	SV_{global}(u) = \sum_{FG \in \{FG\}_{u}}SV(u, FG)
\end{equation}

\subsection{Structural virality and influence--degree}
\label{subsec:virality}

Structural virality is a method for evaluating how viral a retweet cascade tree is \cite{Goel2015:virality}. The structural virality of a cascade tree $T$ with $n>1$ nodes is the average distance between all pairs of nodes in a cascade. That is:
\begin{equation}\label{eq:virality}
	\nu(T) = \frac{1}{n(n-1)}\sum_{i=1}^{n}\sum_{j=1}^{n}d_{ij}
\end{equation}
\noindent where $d_{ij}$ is the shortest path between the nodes $i$ and $j$. The $\nu(T)$ represents the average depth of nodes when we consider all nodes as the root of the cascade. 

We expect that the tree of a viral cascade will have many sub--trees, which represent many generations of a viral diffusion process in a smaller scale. On the other hand, a cascade tree with many leaves, directly connected with the root, represents a ``broadcast'' -- where in a single diffusion process the material has been transmitted to many nodes (see Figure~\ref{fig:example_graph}(a) an example of a broadcast). Even though the structural virality is a measure for the cascade tree, it also reflects the collective influence of the nodes in the tree, meaning that not only the root but also other intermediate nodes should have been influential, since the material has been transmitted in several regions of the network. So, we expect to find influential nodes in cascades with large structural virality. 
Hence, in order to measure the influence on individual level, we define the \textit{influence--degree}.
The influence--degree measures the direct influence a node had on a cascade tree. It is defined as the number of users that have been influenced by a given user $i$ in the cascade tree. For instance, in Figure~\ref{fig:example_graph}(c) the influence--degree of node $a$ is 2 because he has influenced both $b$ and $e$.

The global \textit{influence--degree} is the total number of users that have been influenced by $i$ in all the cascade trees that $i$ has participated in. 

\begin{figure*}[t]
		\centering
		\subfloat[]{\includegraphics[width=.4\linewidth]{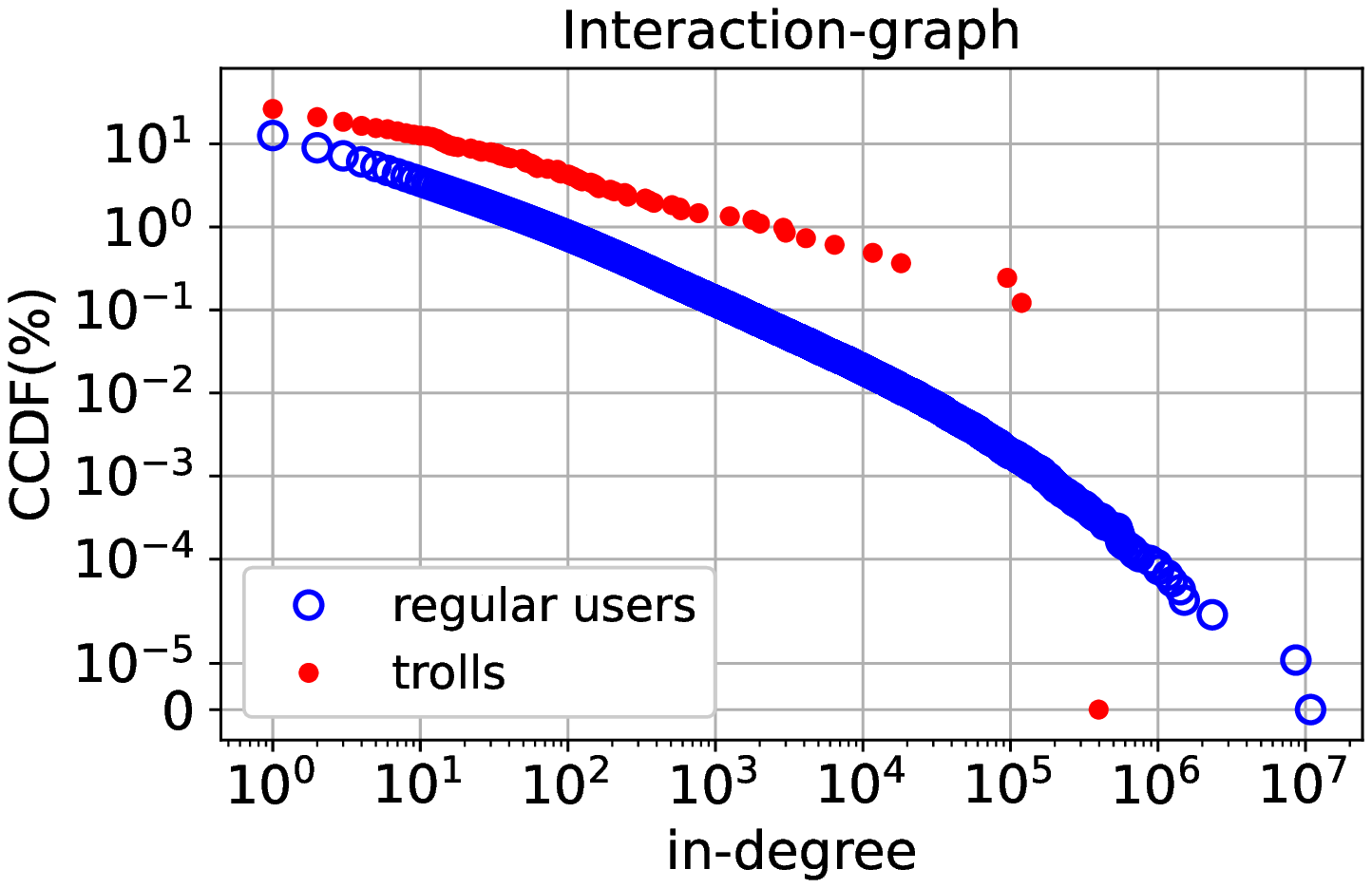}}
        \hfil
		\subfloat[]{\includegraphics[width=.4\linewidth]{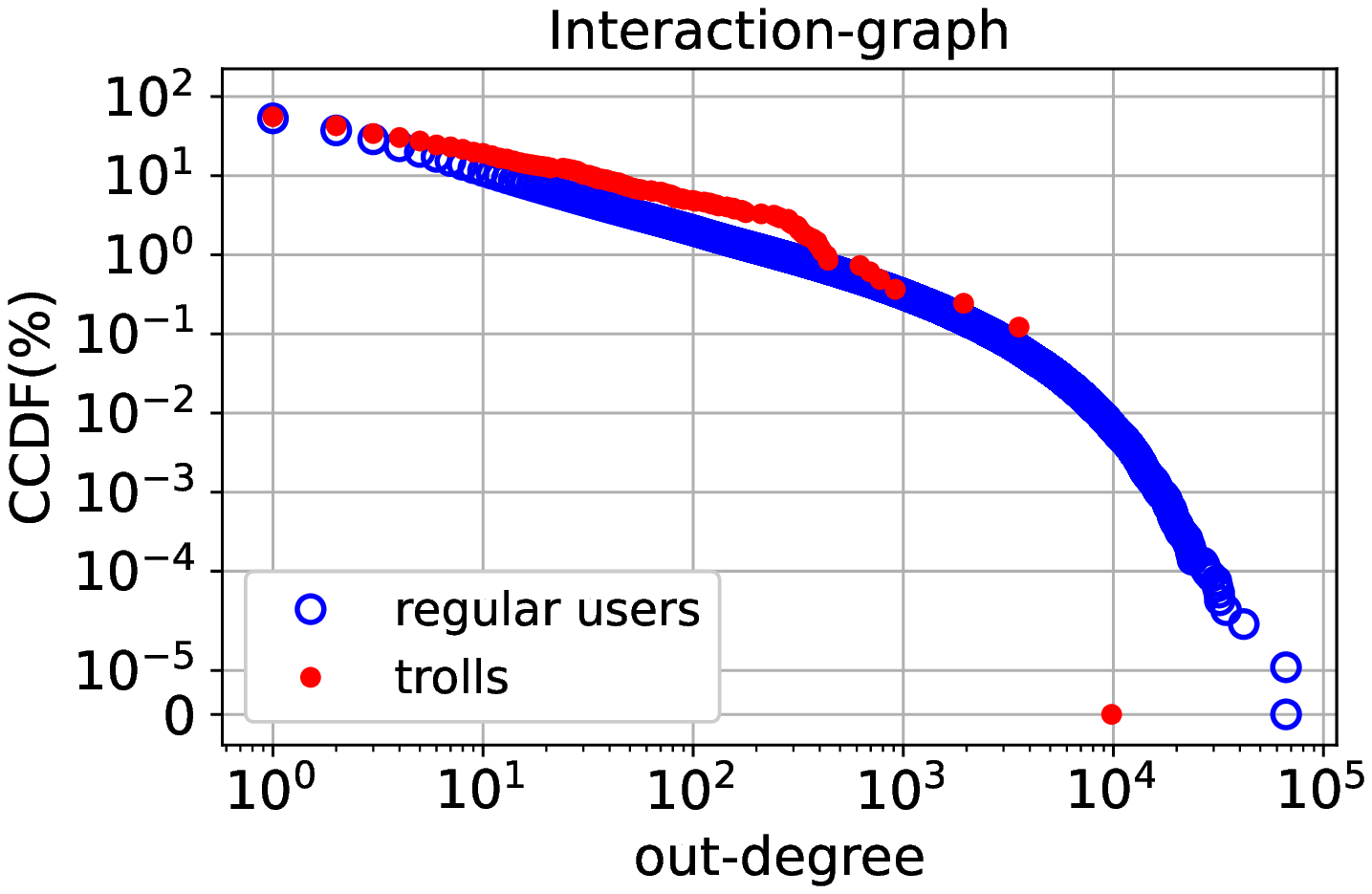}}
		\hfil
		\subfloat[]{\includegraphics[width=.4\linewidth]{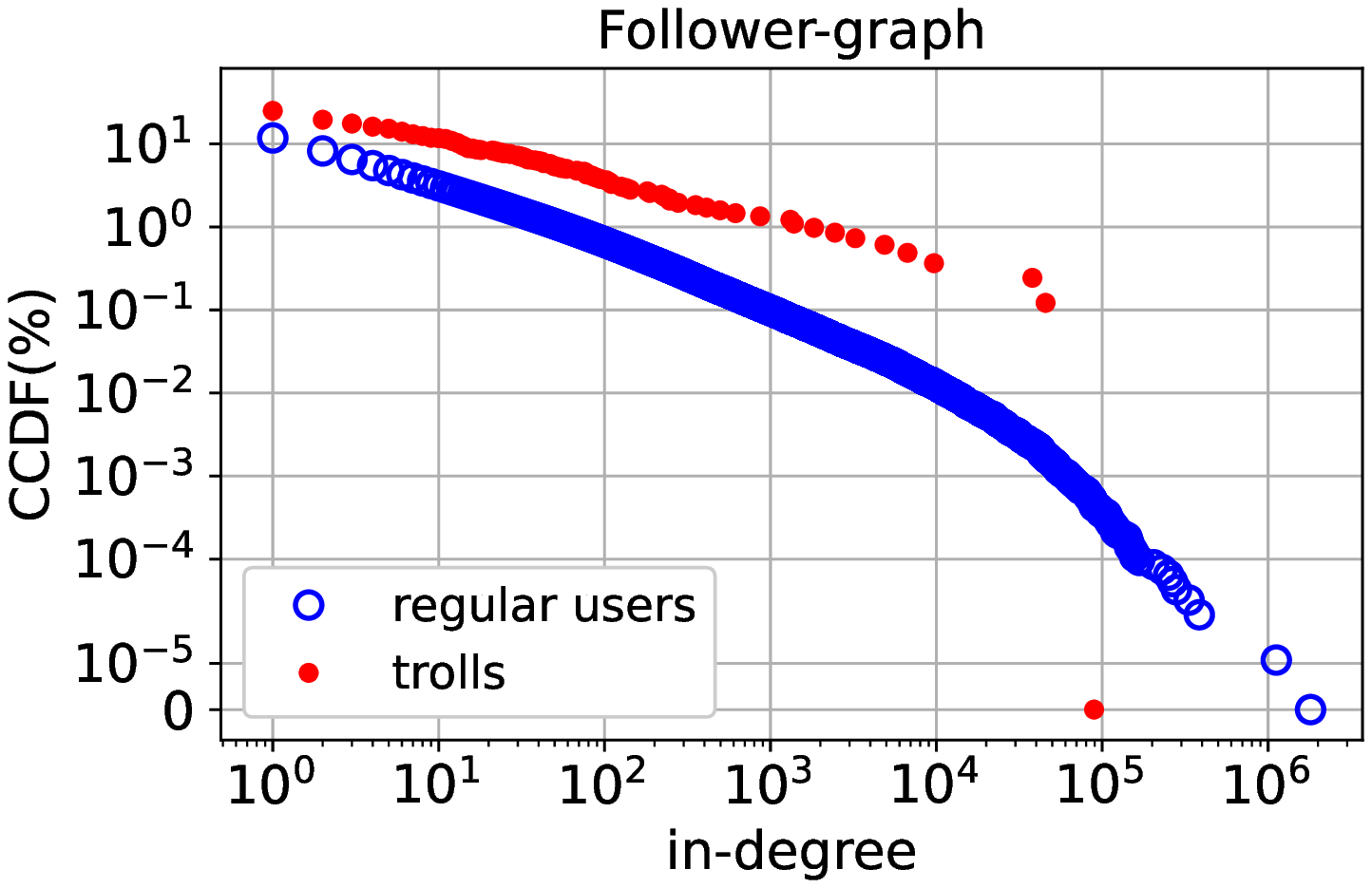}}
        \hfil
		\subfloat[]{\includegraphics[width=.4\linewidth]{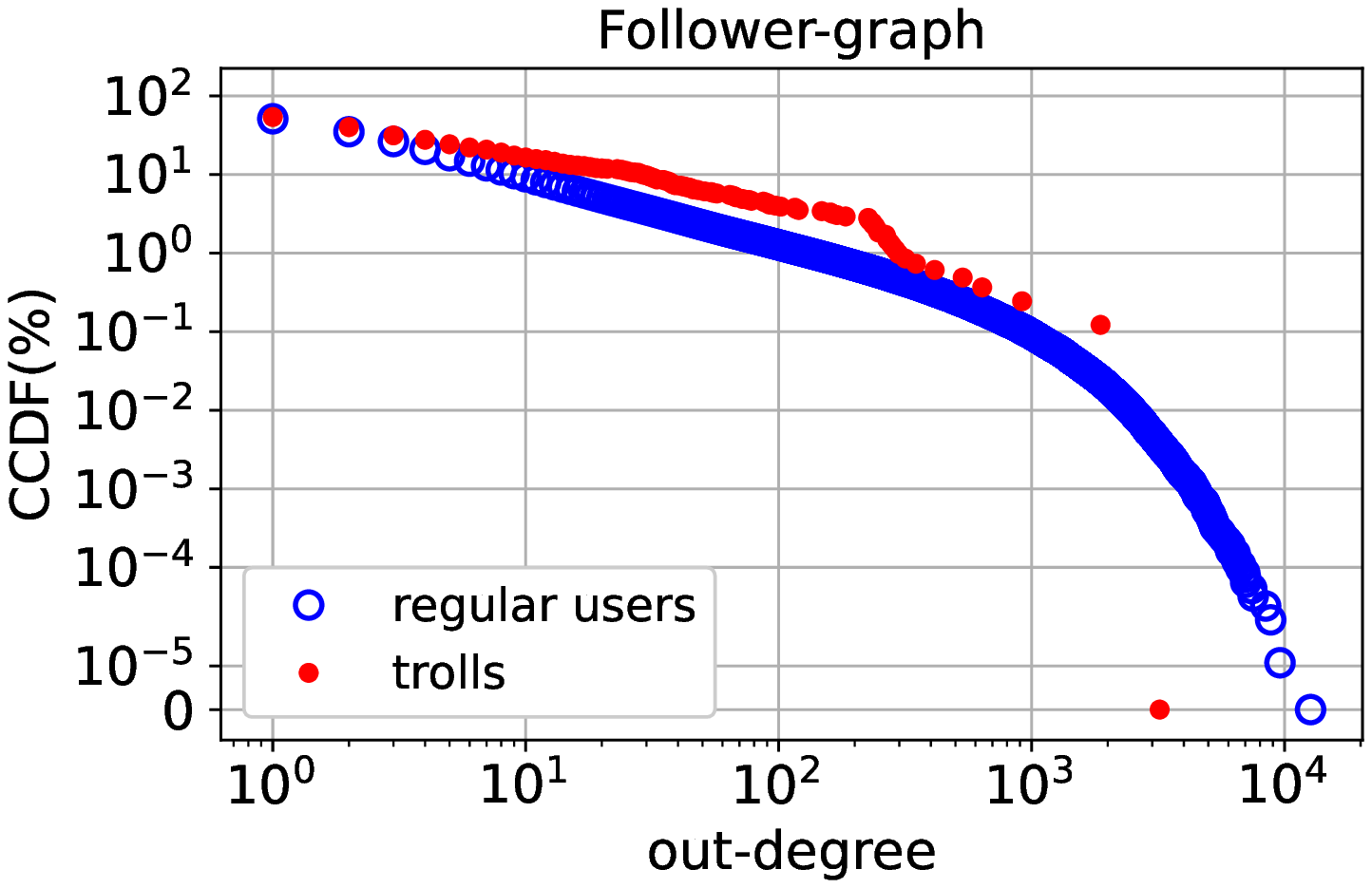}}		
	\caption{CCDF of the non--zero in--degree and out--degree of trolls and regular users.}
	\label{fig:degree}
\end{figure*}

\section{Results}
\label{sec:results}

The analysis is based on the comparison of the influence of two groups of users; the trolls and the regular users. First, we provide general topological features of the interaction--graph, as well as the follower--graph. Next, we focus on the retweet cascades. We compute the users' Shapley Value and influence--degree along with the Structural Virality of the cascade trees. Finally, we provide global rankings where we identify the top--k influential users.

\subsection{Graph topology}
\label{subsec:topology}

\subsubsection{Degree distribution}
\label{subsubsec:degree}

\begin{table}[t]
	\centering
	\caption{Average values: Regular Users vs Trolls}\label{tab:average}
	\setlength\tabcolsep{2pt}
	\begin{tabular}{l  l  c  c}
		&     & Regular Users     & Trolls \\
		\hline
		\multirow{ 2}{*}{Interaction--graph} & In-degree & 18.16  & 821.22 \\
		& Out-degree &  18.23 & 38.97 \\
		\hline           
		\multirow{ 2}{*}{Follower--graph} & In--degree & 8.99  & 258.63 \\
		& Out--degree & 9.02  & 22.48 \\
		\hline
		Largest Comp.& Coreness    &  9.22 &  31.75 \\
		\hline 
		\multirow{ 3}{*}{RT Cascades} & Shapley Value & 3.21  & 269.02 \\
		& Infl. Degree     & 5.35  & 382.71 \\ 
		& Ranking by Shapley          & $1,82\cdot10^{6}$ & $1,61\cdot10^{6}$\\ 
		\hline
	\end{tabular}	                                  
\end{table} 

Figure~\ref{fig:degree} presents the empirical complementary cumulative distribution (CCDF) of the in--degree and out--degree for regular users and trolls in the interaction--graph and follower--graph. The in--degree represents the user's popularity, i.e. the overall activity by his followers on his posts. On the other hand, the out--degree is a measure of a user's sociability/extroversion, i.e. how active a given user is by interacting with other Twitter accounts. Finally, it is important to compare the degree distributions of both interaction--graph and follower--graph. Users with a high degree in the interaction--graph do not necessarily have a large degree in the follower--graph; for instance, users who are highly popular in a small group of followers. Generally, the degree distributions for both graphs are very similar, thus we discuss the findings only for the interaction--graph, since it depicts the overall users' activity.

\begin{figure*}[t]	
	\centering
	\subfloat[]{\includegraphics[width=.4\linewidth]{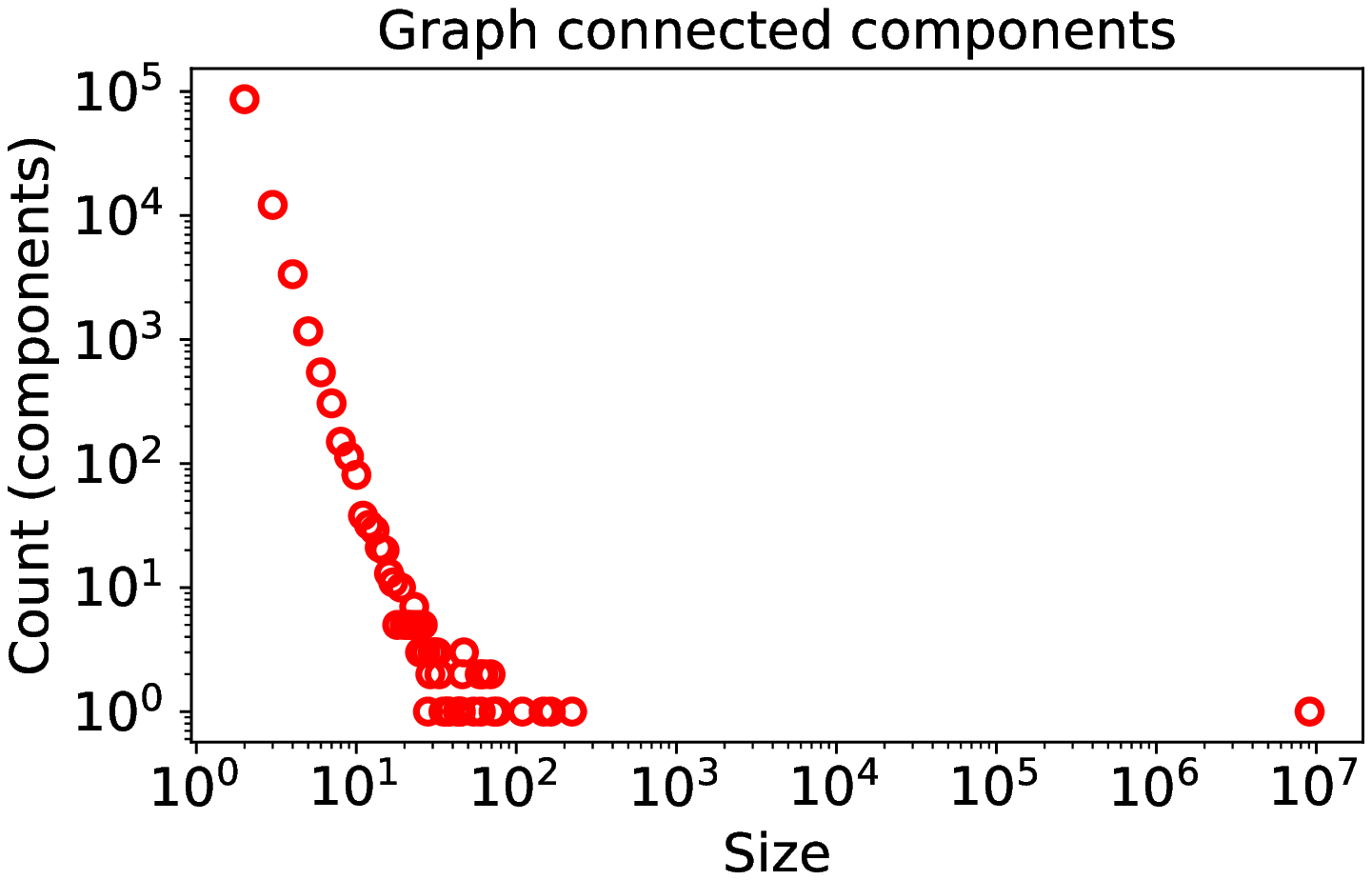}}  
	\hfil
	\subfloat[]{\includegraphics[width=.4\linewidth]{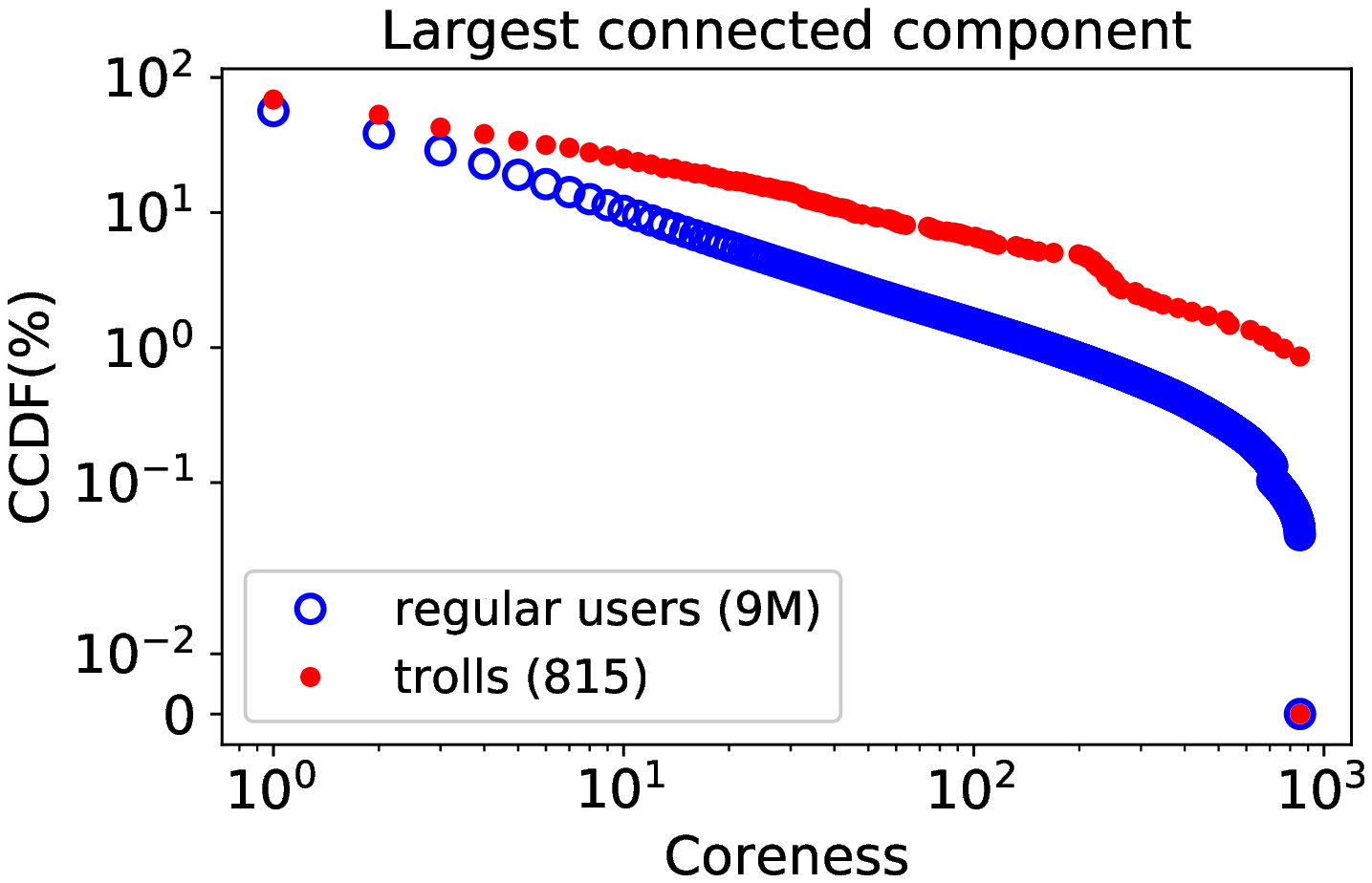}}  
	\caption{(a) The connected components of the undirected version of the follower--graph; (b) CCDF of the coreness for the nodes in the largest connected component.}
	\label{fig:Componets_and_kcore}
\end{figure*}

From Figure~\ref{fig:degree} we observe that: (i) 285 trolls and 2.3M regular users have non zero in--degree; (ii) 675 trolls and 8.5M regular users have non--zero out--degree.

Regarding the in--degree (Figure~\ref{fig:degree}(a)): (i) 12 troll accounts have in--degree larger than 1K. The top--3 trolls have in--degrees 95K, 119K and 396K. One the other hand, we have 12K and 1.8K regular users with in--degree larger than 1K and 10K, respectively. The top--3 regular users have in--degrees 2.3M, 8.6M and 10.8M. 

Regarding the out--degree (Figure~\ref{fig:degree}(b)): (i) the troll activity is not substantial, i.e., 3 accounts have out--degree larger than 1K and the top--3 trolls have out--degrees (1.9K, 3.5K, 9.8K); (ii) the regular users appear to be considerably more active, i.e., 29.6K and 594 accounts have out--degree larger than 1K and 10K, respectively. In conclusion, it seems that in our dataset the troll activity is not dominant compared to the activity of regular users.

Finally, Table~\ref{tab:average} presents the average values for in--degree and out--degree for trolls and regular users in interaction--graph and follower--graph. Even though the regular users are the dominant part of the population, the trolls attracted, on average, a considerably large amount of traffic. For instance, the trolls' average in--degree is 45 times higher than the regular users' average in--degree. 

\subsubsection{Connected components}
\label{subsubsec:components}

Here we examine the structure of the follower--graph by identifying the connected components. Since we have only a sample of the total activity, we examine the undirected version where all the edges (social relationships) are reciprocal. A connected component is a subgraph where for every pair of nodes $i$, $j$ there is an undirected path -- a graph traverse -- from $i$ to $j$. The connectivity of a region implies that there is a possible path for information flow between the nodes that belong to this region.

The undirected graph consists of 9.32M nodes and 82.8M reciprocal edges. We identify 104,954 connected components. Figure~\ref{fig:Componets_and_kcore}(a) presents the number of connected components for a given component size (i.e., number of nodes in the component) in a log--log plot. The largest part of the graph is well--connected. The largest connected component consists of 9M nodes and 82,7M edges while the second largest only has 223 nodes. In other words, we have a giant connected component along with thousands very small ones. 

\subsubsection{k--core decomposition}
\label{subsubsec:kcore}

We compute the \textit{$k$--core decomposition} of the nodes in the largest connected component. The $k$--core decomposition is the process of computing the \text{cores} of a graph $G$. The $k$-core is the maximal subgraph of $G$ where each node has degree at least $k$. The $k$--shell is the subgraph of $G$ that consists of the nodes that belong to $k$--core but not to $(k+1)$--core. A node has \textit{coreness} (or core number) $k$ if it belongs to the $k$--shell. In other words, each node is assigned to a shell layer of the graph $G$. The graph $k$--core number is the maximum value of $k$ where the $k$--core is not empty. It has been proved that the coreness is one of the most effective centrality measures for identifying the influential nodes in a complex network \cite{kitsak:2010}.

In Figure~\ref{fig:Componets_and_kcore}(b), we present the empirical complementary cumulative distribution (CCDF) of the coreness values for trolls and regular users. The graph $k$--core number is 854. The majority of nodes in the larger $k$--shells are the users, since their population is larger than that of the troll accounts. There are only eight trolls with large coreness; seven accounts are part of the largest 854--shell and one account is part of the second largest 853--shell. This is an indication that these accounts were probably influential. Regarding the regular users, 3,710 and 250 of them belong to the largest and second largest k--shell, respectively. Finally, from Table~\ref{tab:average} we observe that the average coreness of trolls is three times larger than the coreness of regular users. 

\noindent\textbf{Summary of Results:} Few trolls have substantial number of followers (in--degree), activity on other accounts (out--degree) and structural position in the network (coreness). Generally, the dominant part of the population is the regular users. On the other hand, on average, the trolls attracted tens of times more traffic than the regular users.

\begin{figure*}[t]	
		\centering
		\subfloat[]{\includegraphics[width=.4\linewidth]{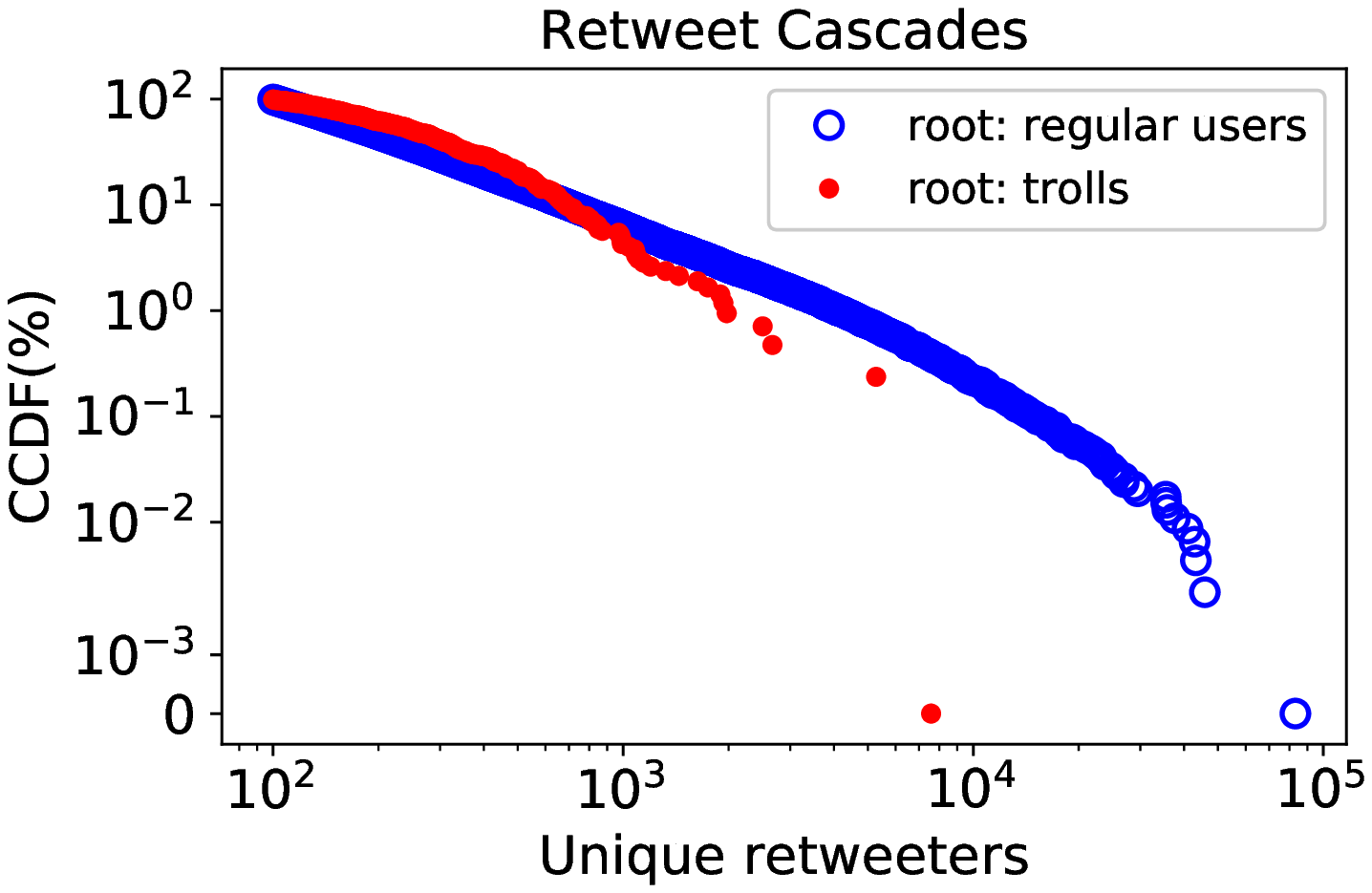}}  
        \hfil
		\subfloat[]{\includegraphics[width=.4\linewidth]{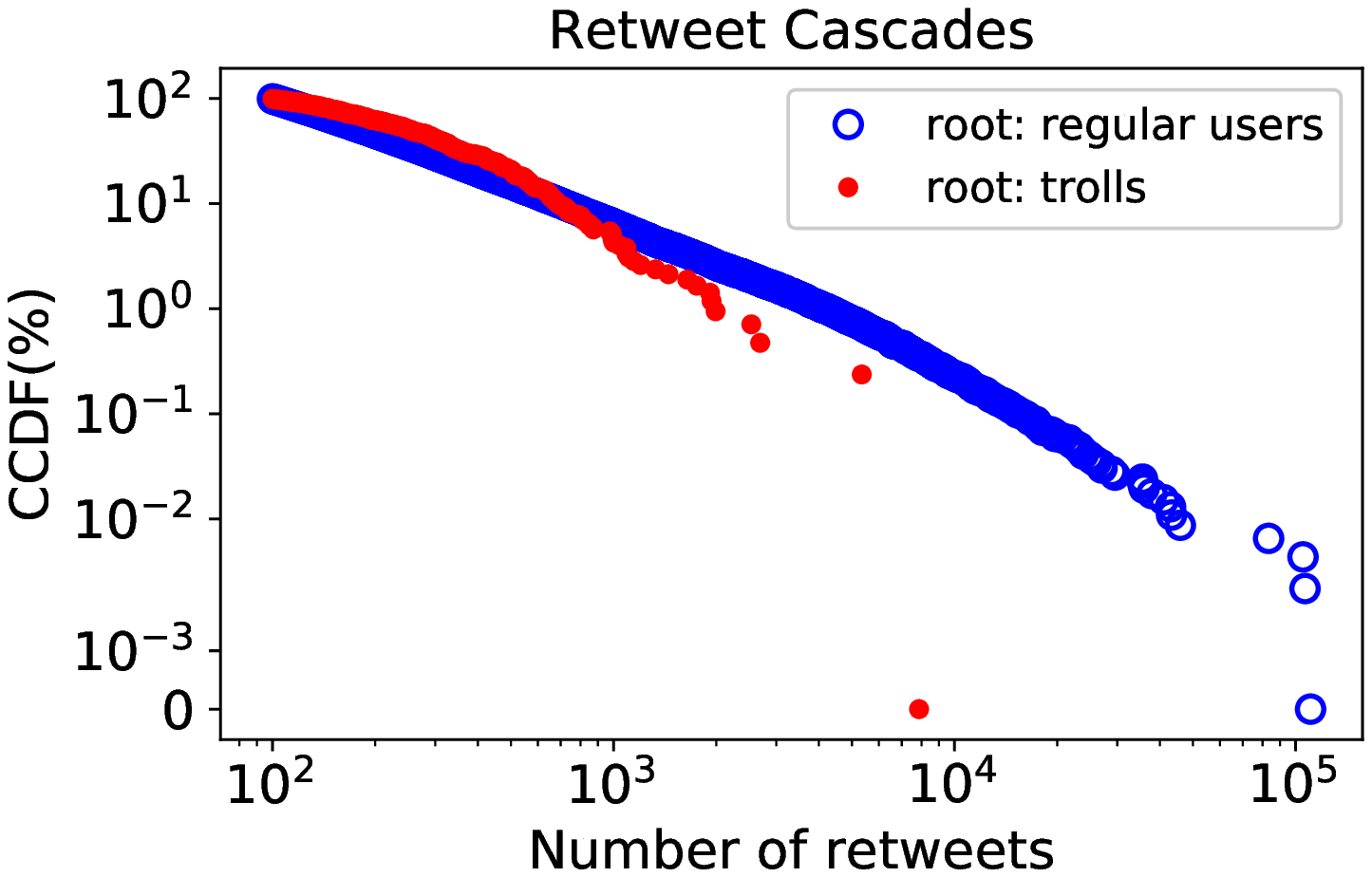}}  
	\caption{CCDF of retweet cascades in terms of unique number of retweeters and total number of retweets. The retweeters might have retweeted the same tweet more than once, hence the number of retweets is larger than the number of retweeters.}
	\label{fig:RTcascades}
\end{figure*}

\begin{figure}[t]	
		\centering
		\includegraphics[width=.8\linewidth]{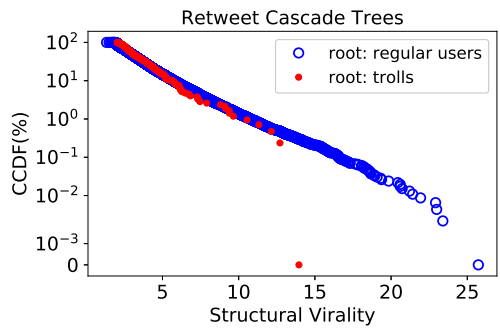}  
	\caption{Structural Virality of the retweet cascade trees.}
	\label{fig:virality}
\end{figure}

\begin{figure*}[t]	
		\centering
		\subfloat[]{\includegraphics[width=.4\linewidth]{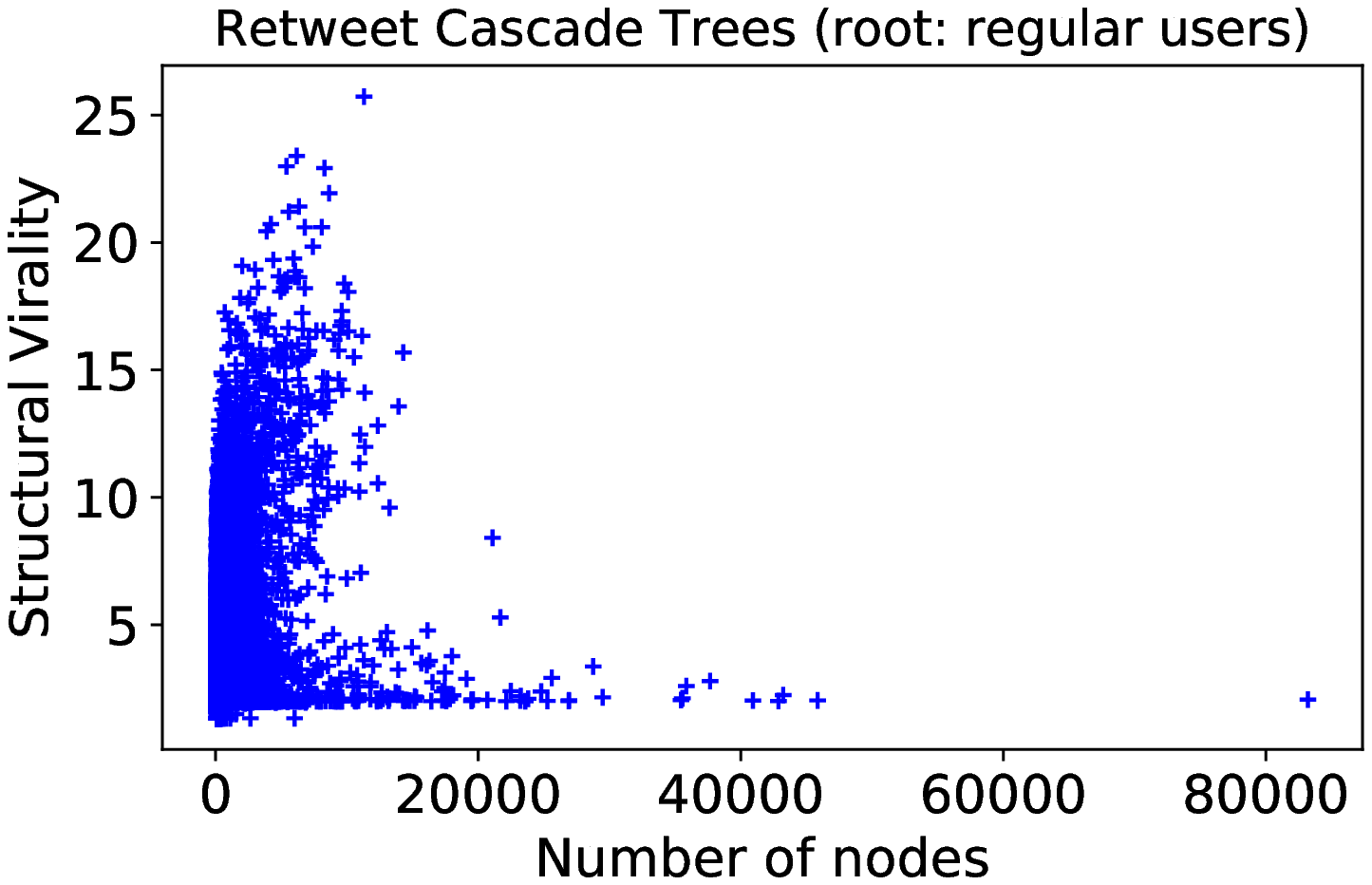}}  
       \hfil
		\subfloat[]{\includegraphics[width=.4\linewidth]{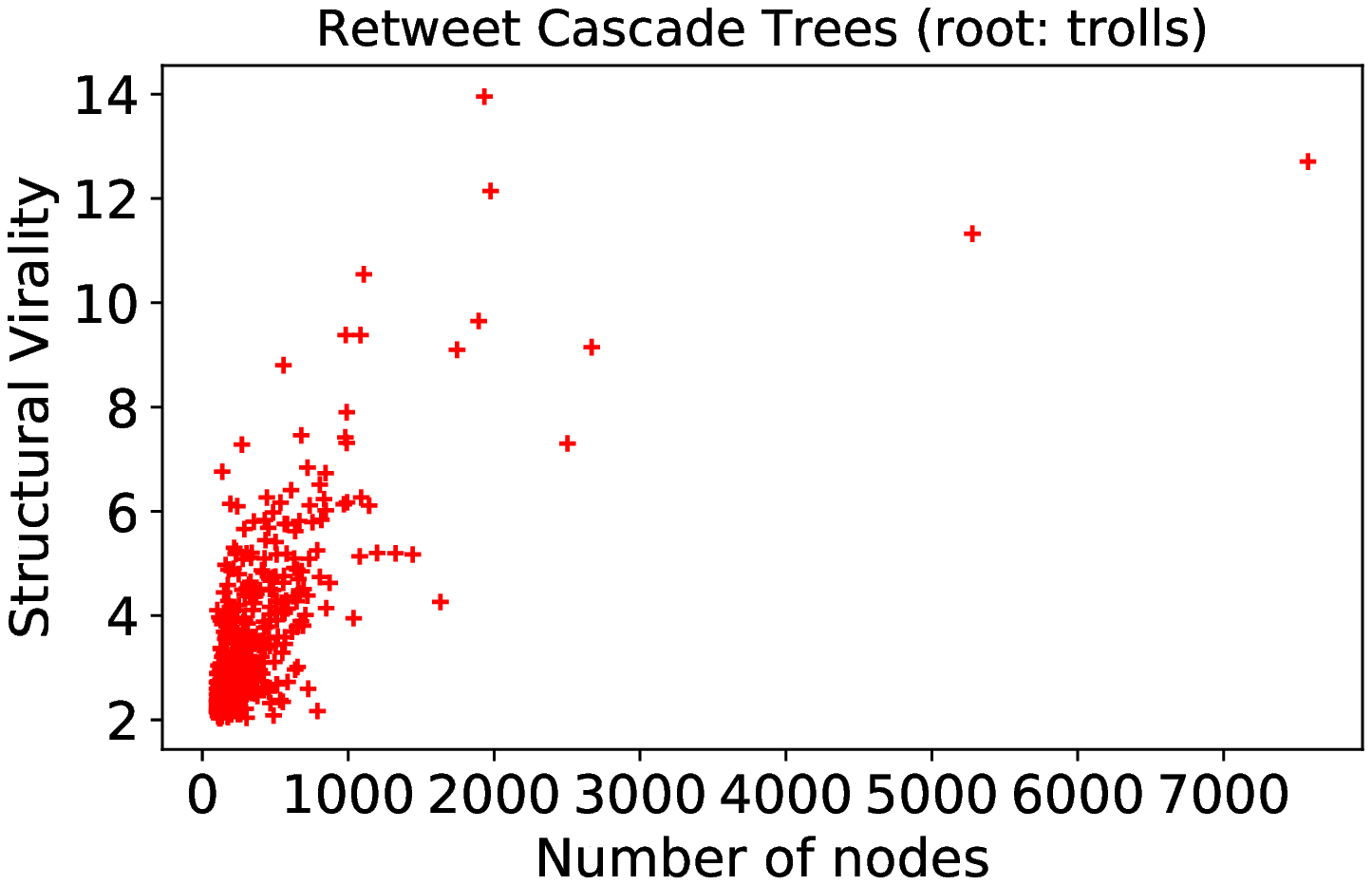}}  
	\caption{Scatter plot of structural virality versus number of retweeters for retweet trees which have either trolls or regular users as root (the author of the original tweet).}
	\label{fig:virality_vs_users}
\end{figure*}

\subsection{Retweet Cascades}
\label{subsec:results_RTcascades}

We now turn our attention to the retweet cascades and we provide general statistics about the popularity of the root tweets posted by regular users and trolls. In Figure~\ref{fig:RTcascades} we present the CCDF of the number of unique retweeters and the CCDF of the total number of retweets per retweet cascade. From the 423 retweet cascades that have been initiated by troll accounts, 18 of them have more than 1K retweeters. In addition, the two largest cascades have 5.2K and 7.5K retweeters (Figure~\ref{fig:RTcascades}(a)). Regarding the cascades that were initiated by regular users, in 2,890 of them the number of retweeters is larger than 1K; 101 cascades have more than 10K retweeters and the top--5 have between 40K to 83.2K. Regarding the number of retweets per cascade, the findings are similar to the previous ones. The most popular root tweets have been posted by regular users instead of trolls (Figure~\ref{fig:RTcascades}(b)). Moreover, in the largest four cascades, the number of retweets is between 83K to 111K, which renders them considerably larger than the number of unique retweeters. This indicates that the root tweets of these four cascades were very popular and they have been retweeted multiple times by the same users.

\subsubsection{Structural virality}
\label{subsubsec:results_virality}

The previous results depict that the cascades initiated by trolls were not considerably large. However, the results are based on the unstructured raw data provided by Twitter API, where all the retweets point to the original tweet (see the example in Figure~\ref{fig:example_graph}(a)). Here, we aim to measure how viral the cascades were by using the measure of structural virality (see Section~\ref{subsec:virality}). For the computation of Equation~\ref{eq:virality}, we use the networkx\footnote{\url{https://networkx.github.io/}} Python package (Dijkstra's algorithm). 

In Figure~\ref{fig:virality} we compare the structural virality of cascade trees for: (i) the cascades initiated by trolls (423 root--tweets, see Table~\ref{tab:retweets}); and (ii) the 45,986 cascades initiated by regular users. We can see that the regular users were the source of the most viral cascades. The top troll cascade has 13.95 structural virality. On the other hand, 138 user cascades have structural virality larger than 13.95.

One would expect that cascades with large structural virality, should also have large number of participants (retweeters). In other words, we should expect a positive correlation between these two variables. However, this does not seem to be the case in our dataset. Specifically, in Figure~\ref{fig:virality_vs_users} we present the scatter plots of structural virality versus the number of nodes in the cascade tree. We observe that cascades with very small virality have a quite large number of users (Figure~\ref{fig:virality_vs_users}(a)). This means that the majority of users retweeted the original tweet and not so often the retweet of another user. For the cascades that were initiated by troll accounts, the situation is different (Figure~\ref{fig:virality_vs_users}(b)). There are cascades with very large virality and a large number of users (regular or troll). A possible explanation is that the community around the trolls was more dense, with users retweeting each other and forming an ``echo chamber'' where political polarization took place. 

\noindent\textbf{Summary of Results:} The vast majority of viral cascades were initiated by regular users and very few by troll accounts. Moreover, retweet cascades with thousands of retweets have very small structural virality, which indicates that their root users were the main source of influence.
 
\subsection{Top--k influential users}
\label{subsec:topk_users}

We conclude the analysis by identifying the most influential Twitter accounts. We estimate their influence based on two measures, the Shapley Value--based centrality and the influence--degree. First, we produce the global ranking of all accounts that are part of the retweet cascades (see Table~\ref{tab:retweets}; 233 trolls; 3.6M regular users). Then we report the rank of the most influential trolls and regular users. In addition, we measure how close to a Twitter bot the profiles of the top--1000 regular users are. In order to estimate this we use the Botometer API \cite{botometer:varol:2017,botometer:yang2019scalable} which has been used in the literature for identifying Twitter bots~\cite{Bovet2018,Bovet:2019}. Our goal is to examine whether the behavior of top ranked accounts deviate from a human--operated account. As we mentioned in Section~\ref{sec:intro}, an account can be automated (having a high Botometer score) and at the same time can be benign. Hence, being a Twitter bot does not coincide with being a troll. On the other hand, a high bot--score raises questions about the authenticity of the account. Finally, we collect the top--1000 accounts information and we identify the suspended accounts. 

\begin{figure*}[t]
		\centering
		\subfloat[]{\includegraphics[width=.4\linewidth]{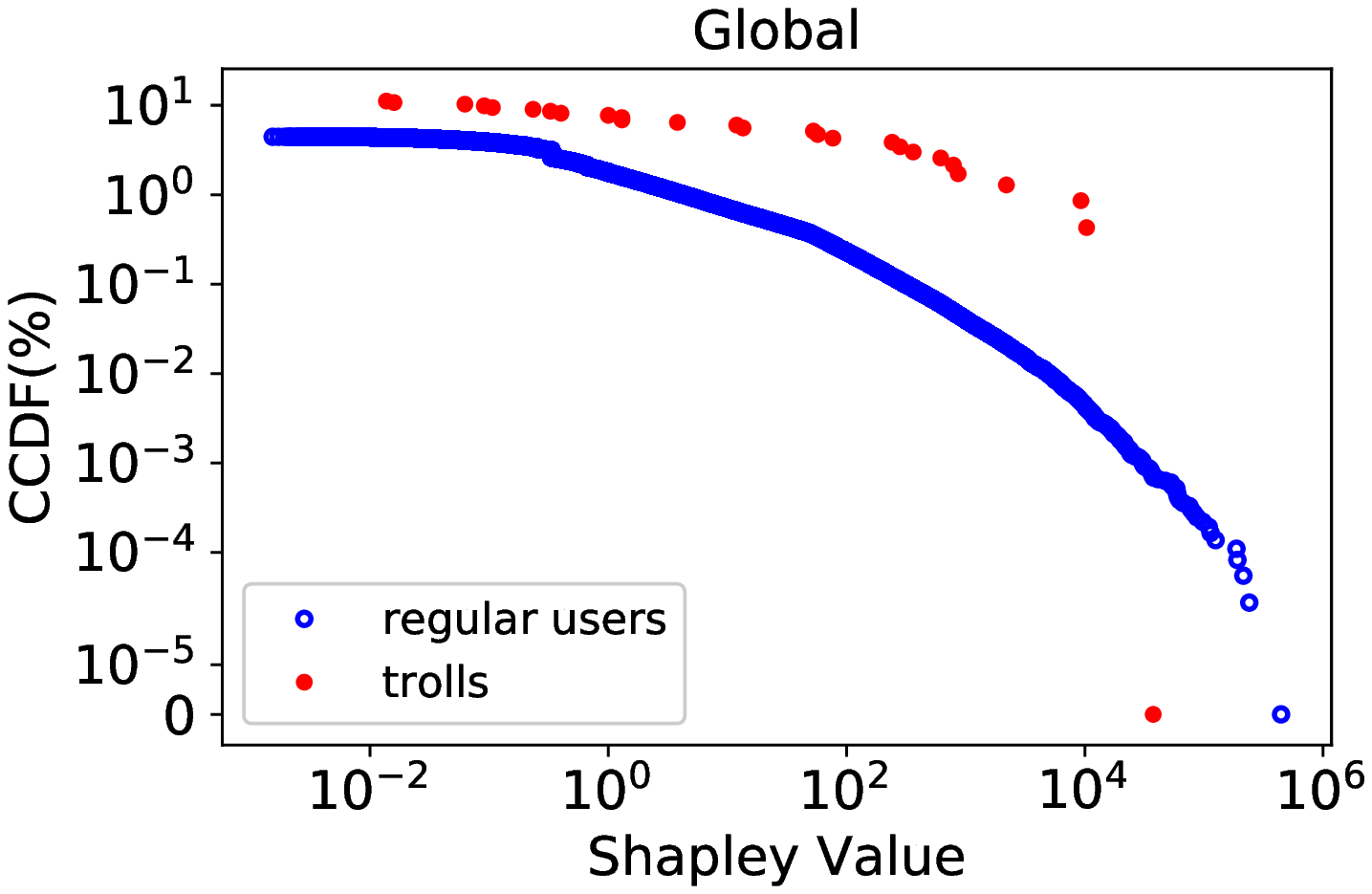}}  
        \hfil
		\subfloat[]{\includegraphics[width=.4\linewidth]{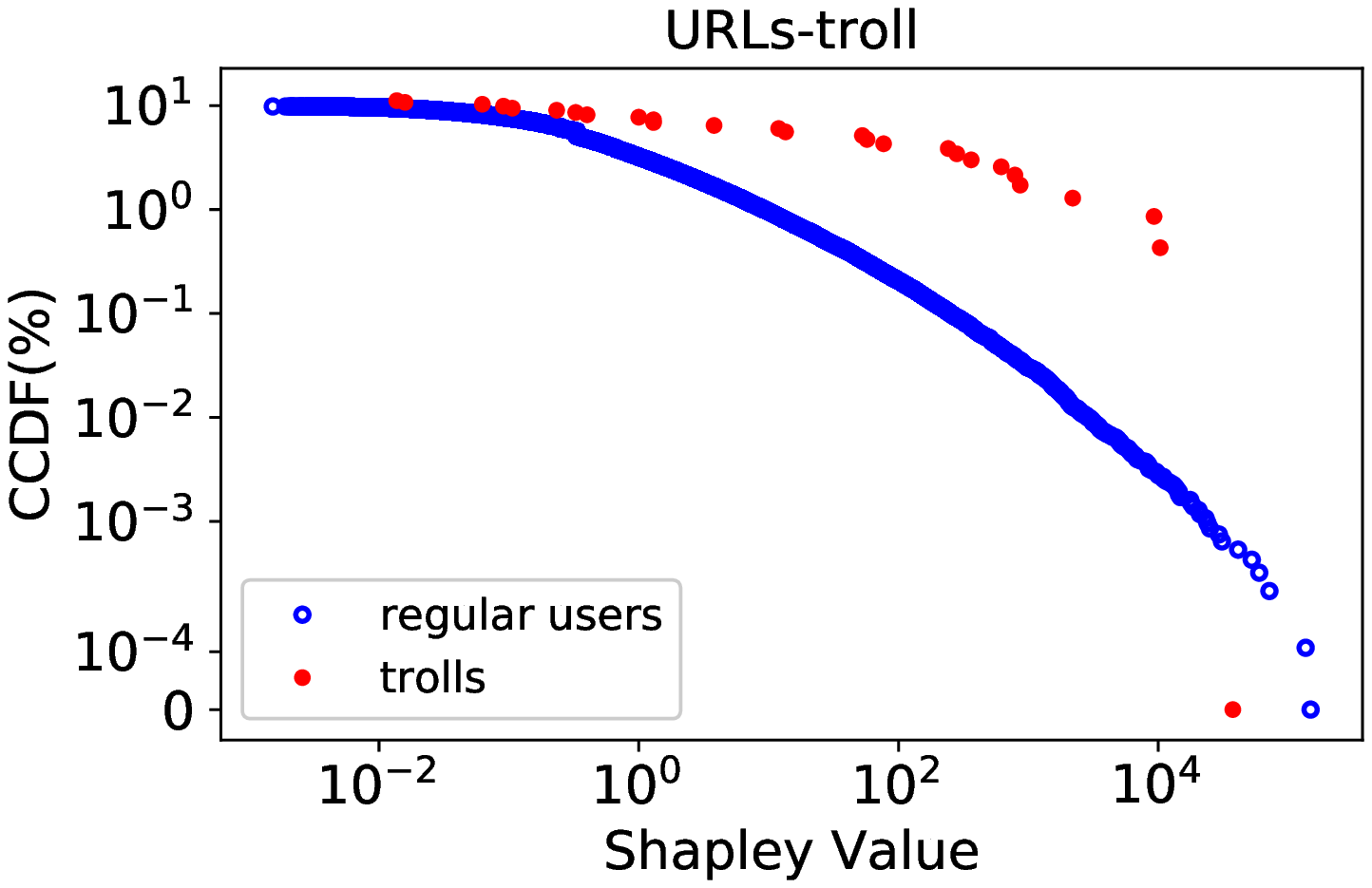}}  
	\caption{CCDF of the non-zero Shapley Value for trolls and regular users.}
	\label{fig:shapley}
\end{figure*} 

\begin{table*}[!t]
	\setlength\tabcolsep{3pt}
	\caption{Top--k accounts}
	\label{tab:topk}
    \centering
	\begin{tabular}{l l  c c  c  c c }
		\hline
		User ID & Screen--name & Ranking     &  Ranking      & Coreness & CAP(English) & CAP(Universal) \\
		&              &  by Shapley V.         &  by Infl. Degree &         &              &              \\              
		\hline
		\multicolumn{7}{c}{\textbf{Top--10 influential accounts}} \\		
		1339835893         &   HillaryClinton &                1 &         1 &   854 &  0.00148281 &    0.00193585 \\
		347627434          &      LindaSuhler &                2 &         2 &   854 &  0.00682617 &     0.0185992 \\
		25073877           &  realDonaldTrump &                3 &         4 &   854 &  0.00155435 &    0.00219844 \\
		729676086632656900 &        TeamTrump &                4 &         5 &   854 &  0.00141879 &    0.00186551 \\
		16589206           &        wikileaks &                5 &         6 &   854 &  0.00126169 &    0.00186551 \\
		\textbf{779739206339928064} & \textbf{WDFx2EU7 (Suspended)} &                6 &        13 &   854 &         N/A &  N/A \\
		18643437           &     PrisonPlanet &                7 &         9 &   854 &   0.0011786 &    0.00201394 \\
		1367531            &          FoxNews &                8 &         7 &   854 &  0.00281466 &    0.00257104 \\
		\textbf{431917957 } &  \textbf{magnifier661 (Suspended)} &                9 &        11 &   854 &         N/A &  N/A \\		
		759251             &              CNN &               10 &         8 &   854 &  0.00313413 &    0.00272963 \\
		\textbf{732980827} & \textbf{ChristiChat (Suspended)}   &               11 &        10 &   854 &       N/A &         N/A \\
		\textbf{355355420} & \textbf{StylishRentals (Suspended)}  &               13 &        3  &    96 &       N/A   &    N/A     \\
		\hline
		\multicolumn{7}{c}{\textbf{Potentially Bot accounts in top--1000}} \\
		3248410062 &         rsultzba &              275 &                412 &   854 &    0.565053 &      0.296769 \\
		717627639159128064 &    TrumpLadyFran &              311 &                355 &   854 &    0.846935 &       0.44586 \\
		429229693 &        edeblazim &              531 &                571 &   854 &    0.892355 &      0.812234 \\
		2913627307 &  WORIDSTARHIPHOP &              643 &                552 &    46 &    0.511452 &      0.385459 \\		
		\hline
		\hline
		User ID & Screen--name &  Ranking             &  Ranking             & Coreness & \multicolumn{2}{c}{\#Retweeted}\\
		&              &  by Shapley V.     &  by Infl. degree     &   & by top--1000 Users & by top--10 Users\\     
		\hline
		\multicolumn{7}{c}{\textbf{Troll accounts in top--1000}} \\		
		4224729994 &    \verb|TEN_GOP|        &   27 &    34  &   854    &  1098 (1.11\%) & 17 \\
		4272870988 &   \verb|Pamela_Moore13|  &   150 &   201 &   854    &  372 (1.32\%) & 4 \\
		4218156466 &   \verb|America_1st_|    &   181 &   241 &   854    & 347  (1.34\%) & 3 \\
		3990577513 &    tpartynews            &   769 &   899 &   854    &  63  (1.15\%) & 2 \\
		\hline							
	\end{tabular}
\end{table*} 

\subsubsection{Shapley Value--based centrality}
\label{subsubsec:results_shapley}

As we mentioned in\\ Section~\ref{subsec:flow-graph-tree}, for each retweet cascade there is a corresponding flow graph which formalizes the overall information exchange that has taken place between the retweeters. Here, based on the flow graphs and the Equations~\ref{eq:shapley} and~\ref{eq:global_shapley}, we compute the global Shapley Value of every user who participated in the retweet cascades. Moreover, having the URLs that are embedded in the root-tweets as identifiers of the web and media material that has been diffused in the network, we collect only the cascades that refer to URLs that have been spread by trolls -- either by posting an original root tweet or by retweeting. For simplicity, we call these URLs as \textit{URLs--troll}. 

In Figure~\ref{fig:shapley}(a) we plot the CCDF of the global Shapley Values. We have 27 out of 233 trolls and 161,513 out of 3.6 million regular users with non--zero Shapley Value. In other words, only 27 trolls have non-zero contribution on the diffusion of information that took place by the retweet cascades. Subsequently, based on the global Shapley Values, we get the global ranking, where the rank for the trolls is [27, 150, 181, 769, 1649, 1797, 2202, 3273, 3964, 4424, 10017, 12263, 12939, 22706, 23858, 38246, 58516, 58524, 64181, 90589, 114414, 124387, 139794, 142181, 146944, 158378, 158960].\\ Hence, only four troll accounts are in the top--1000 and one of them in the top--100 (see Table~\ref{tab:topk}). Moreover, from Table~\ref{tab:average} we observe that the average ranking of trolls is not significantly larger than the rest of the population. At the same time, the average Shapley Value (global) for troll accounts is 83 times larger than the regular users' Shapley Value, which indicates that the trolls accounts were quite effective in spreading information.

Finally, in Figure~\ref{fig:shapley}(b) we report the Shapley Values only for the retweet cascades of \textit{URLs--troll}. We have 2,723 URLs which appear in 3,924 cascades consisting of 934K regular users and 233 trolls, in total. 27 trolls and 91,572 regular users have non--zero Shapley Value. The distribution for the trolls is the same with the global one, since the retweet cascades of \textit{URLs--troll} are the only ones with troll accounts present. Regarding the regular users, we recompute their total Shapley Value by the Equation~\ref{eq:global_shapley} and only for the subset of retweet cascades that correspond to \textit{URLs--troll}. Again, we reach a final ranking, where the ranking of trolls in the top--1000 is [7, 28, 32, 125, 335, 361, 444, 697, 864, 981], namely, ten trolls appear in the top--1000 and three of them in the top--100.

\subsubsection{Influence--degree}
\label{subsubsec:results_influence}

\begin{figure}[!t]
		\centering
		\includegraphics[width=.8\linewidth]{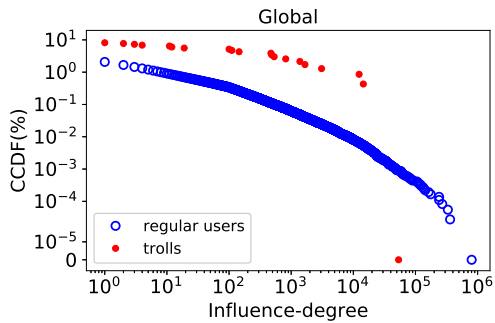}  
	\caption{CCDF of the non-zero Influence--degree for trolls and regular users.}
	\label{fig:influence}
\end{figure}

Now, we use the influence--degree as a measure to rank regular users and trolls according to the effect they have on the retweets cascade trees. Recall that the influence--degree of a given user $i$ is the total number of users that have been directly influenced by $i$ (for more details see Section~\ref{subsec:virality}). Figure~\ref{fig:influence} reports the CCDF of non-zero influence values for regular users and trolls. In summary, we have 21 trolls and 118,960 regular users with non--zero influence. We found four troll accounts in the top--1000 with rankings [34, 201, 241, 899] and one of them in the top--100 (see also Table~\ref{tab:topk}). Finally, from Table~\ref{tab:average} we observe that the influence--degree of trolls is more than 70 times larger than regular users' influence, on average; a similar result with the one for the Shapley Value.

\subsubsection{Bots and suspended accounts}
\label{subsubsec:results_bots}

How similar to bot accounts are the top--k users? In order to estimate this, we use the Botometer scores for the top--1000 regular users (ranking by Shapley Values). Botometer\footnote{\url{https://botometer.iuni.iu.edu}} classifies Twitter accounts as bot or human with 0.95 AUC classification performance~\cite{botometer:varol:2017}. It uses various machine learning models and more than a thousand features which have been extracted from the publicly available data of the account in question, such as friends, accounts' profile or language. When we check an account, the Botometer API returns various scores where the more general one is the \textit{Complete Automation Probability} (CAP) -- the probability that a given account is completely automated. For a given account, two CAP scores are provided, one based on its English language tweets and one for universal features. For instance, if we know that a user is from China and the majority of his/her tweets are written in Chinese, then we can consider the CAP(Universal) score as the estimator of that account being a bot. Generally, the recommendation is that a CAP score above 0.5 indicates a bot account~\cite{Bovet2018}. 

In top--1000, four regular users have either CAP(English) or\\ CAP(Universal) score larger than 0.5, so they are potentially bots (see Table~\ref{tab:topk}). On the other hand, only 22 and 21 users have\\ CAP(English) and CAP(Universal) larger than 0.2. Moreover, 263 accounts were inactive. In fact, during the last few months, we observed that the number of inactive accounts increased from 243 (initially) to 263 (now).

In order to verify the reasons of inactivity, we get the accounts information of the regular users in the top--10000, using Tweepy. When an account is not accessible then the Tweepy returns an error message\footnote{\url{https://developer.twitter.com/en/support/twitter-api/error-troubleshooting}} either "User not found" (code 50; corresponds with HTTP 404; deleted account by the user itself) or "User has been suspended" (code 63; corresponds with HTTP 403; suspended account by Twitter due to violation of Twitter Rules\footnote{\url{https://help.twitter.com/en/managing-your-account/suspended-twitter-accounts}}). In summary, we found that (i) in top--100: 23 suspended accounts out the of 26 inactive ones; (ii) in top--1000: 220 suspended out of the 263 inactive; (iii) in top--10000: 1,836 suspended out of the 2,508 inactive.   

Lastly, Table~\ref{tab:topk} shows the account information for the top--10 influential users based on the Shapley Value. We also present the corresponding rankings in terms of influence--degree and coreness along with the Botometer scores CAP(English, Universal). Two accounts in top--10 are suspended, which raises serious doubts about the authenticity of these users. The top--10 users are part of the largest 854--shell. In addition, we report the four trolls in top--1000 along with their rankings and coreness. All four of them are part of the largest 854--shell. Moreover, in retweet cascades initiated by them, more than 1.1\% of the total number of retweets were from regular users belonging to the top--1000 group.

\noindent\textbf{Summary of Results:} Four troll accounts were amongst the most influential users. Their tweets have been retweeted tens of times by top--1000 influential regular users. Four regular users in the top--1000 exhibit bot behavior. In addition, 23\% and 22\% of regular accounts in the top--100 and top--1000 respectively, have now been suspended by Twitter, something that raises questions about their authenticity and practices, overall.

\section{Conclusion}
\label{sec:conclusions} 

In this paper, we have extensively studied the influence that state--sponsored trolls had during the 2016 US presidential election by analyzing millions of tweets from that period. We first constructed the interaction--graph between trolls and regular users which represent an approximation of the true social network and then we concentrate our analysis on the retweet cascades. In order to measure the users' impact on the diffusion of information, we introduce the notion of \textit{flow graph}, where we apply a game theoretic--based centrality measure. Moreover, we estimate the retweet paths by constructing the retweet cascade trees where we measure the users' direct influence. The results indicate that although the trolls initiated some viral cascades, their role was not a dominant one and the source of influence was mainly the regular users. On the other hand, the average influence of trolls was roughly 70 times more than the influence of the regular users. This indicates that, the strategies these trolls followed in order to attract and engage regular users were sufficiently effective. Furthermore, 23\% and 22\% of regular accounts in the top--100 and top--1000 respectively, have now been suspended by Twitter. This raises questions about the authenticity of these accounts. 

\begin{acks}
We are grateful to Twitter for providing access to the unhansed version of the trolls' ground truth dataset. We thank Nikolaos Laoutaris for his insightful comments about the Shapley Value. This project has received funding from the European Union's Horizon 2020 Research and Innovation program under the Cybersecurity CONCORDIA project (Grant Agreement No. 830927) and the INCOGNITO project (Grant Agreement No. 824015).
\end{acks}

\bibliographystyle{ACM-Reference-Format}
\bibliography{Arxiv_v3}
\end{document}